\pdfoutput=1
\documentclass{JINST}
\let\ifpdf\relax
\usepackage[numbers]{natbib}
\usepackage{graphicx}
\usepackage[latin1]{inputenc}
\usepackage[english]{babel}
\usepackage{placeins}
\usepackage{url}
\usepackage{numprint}
\usepackage{amssymb}
\usepackage{afterpage}
\npdecimalsign{.}
\npthousandthpartsep{} %Scheinbar Absicht dass er auch nach dem Komma Trennzeichen macht, l"asst sich aber beheben.

\newcommand{\np}{\numprint}

\renewcommand{\deg}{^\circ}
\newcommand{\mr}{\mathrm}
\newcommand{\ud}{\mathrm{d}}

\title{Instrumentation for comparing night sky quality and atmospheric conditions of CTA site candidates }%Instrumentation for CTA site characterization}

\author{Christian Fruck$^a$\thanks{Corresponding author.}, 
Markus Gaug$^{b,c}$,%\thanks{Corresponding author.}, 
Jean-Pierre Ernenwein$^d$,
Du\v{s}an Mand{\'a}t$^e$,
Thomas Schweizer$^a$,
Dennis H{\"a}fner$^a$,
Tomasz Bulik$^f$,
Marek Cieslar$^f$,
Heide Costantini$^d$,
Michal Dominik$^f$,
Jan Ebr$^g$,
Markus Garczarczyk$^h$,
Eckart Lorentz\thanks{deceased},
Giovanni Pareschi$^i$,
Miroslav Pech$^e$,
Irene Puerto-Gim{\'e}nez$^j$,
Masahiro Teshima$^a$ \\
\llap{$^a$}Max-Planck-Institut f{\"u}r Physik, F{\"o}hringer Ring 6, 80805 M{\"u}nchen, Germany. \\
\llap{$^b$}F{\'i}sica de les Radiacions, Departament de F{\'i}sica, Universitat Aut{\`o}noma de Barcelona\\ Campus UAB, 08193 Bellaterra, Spain.\\
\llap{$^c$}CERES, Universitat Aut{\`o}noma de Barcelona-IEEC, 08193 Bellaterra, Spain.\\
\llap{$^d$}Aix Marseille Universit{\'e}, CNRS/IN2P3, CPPM UMR 7346, 13288 Marseille, France.\\
\llap{$^e$}Institute of Physics of the Academy of Science of the Czech Republic, Joint Laboratory of Optics, 17. listopadu 12, 771 46 Olomouc, Czech Republic.\\
\llap{$^f$}Astronomical Observatory, University of Warsaw, Aleje Ujazdowskie 4, 00-478 Warsaw, Poland.\\
\llap{$^g$}Institute of Physics of  Academy of Science of the Czech Republic\\ Na Slovance 2, Praha, Czech Republic.\\
\llap{$^h$}Deutsches Elektronen-Synchrotron, Platanenallee 6, Zeuthen, Germany.\\
\llap{$^i$}INAF - Osservatorio Astronomico di Brera, Italy.\\
\llap{$^j$}Institut de Fisica d'Altes Energies (IFAE), E-08193, Barcelona, Spain.\\
E-mail: \email{fruck@mpp.mpg.de, markus.gaug@uab.cat} }
\abstract{%
Many atmospheric and climatic criteria have to be taken into account for the selection of a suitable site for the next generation of imaging air-shower Cherenkov telescopes, 
the ``Cherenkov Telescope Array'' CTA.
Such data are not available with sufficient precision, thus a comparison of the proposed sites and final decision based on a comprehensive characterization is impossible.  
Identical cross-calibrated instruments have been developed which allow for precise comparison between sites, the cross-validation of existing data, and the ground-validation of satellite data.
The site characterization work package of the CTA consortium opted to construct and deploy 9 copies of an autonomous multi-purpose weather sensor, 
incorporating an infrared cloud sensor, a newly developed sensor for measuring the light of the night sky, and an All-Sky-Camera,
the whole referred to as Autonomous Tool for Measuring Observatory Site COnditions PrEcisely (ATMOSCOPE).
We present here the hardware that was combined into the ATMOSCOPE and characterize its performance.
}
\keywords{site selection;atmospheric influence;light of night sky;CTA}

\begin{document}
\section{Introduction}

%%%%%%%%%%%%%%%%%%%%%%

Ground-based observations in the regime of very high energy (VHE) gamma-rays using Cherenkov light produced in extensive air showers constitute a young branch of astronomy. 
The aim of the future Cherenkov Telescope Array (CTA)~\citep{cta}, currently finishing its preparatory phase, is to leave behind the pioneering phase of current Imaging Air-shower 
Cherenkov Telescopes (IACTs)~\citep{hillas2013} 
and provide the astrophysical community with one mature and reliable gamma-ray observatory in both hemispheres, for the observation of gamma-rays with energies 
from a few tens of~GeV to more than 100~TeV~\citep{ctaconcept}. 
This entails a strong requirement to maximize the duty cycle of the observatories and hence minimize the influence of the various atmospheric disturbances, 
as well as possible anthropogenic contributions to the Light of the Night Sky (LoNS), at the main wavelengths covered by the 
Cherenkov light, namely the UV and blue. A very careful selection of the two sites is therefore crucial.

The requirements for an IACT site are different in many ways from those for optical or infrared (IR) telescopes. 
%Since IACTs observe the development of air showers throughout the atmosphere 
%at different heights, the altitude of atmospheric phenomena plays an important role in negotiating energy threshold, collection area and angular resolution. 
%The effective duty cycle, now energy and angular resolution dependent,  is important due to the need for long integration times. 
Since IACTs observe the development of air showers throughout the atmosphere at different heights, the altitude of atmospheric phenomena plays an important role when it comes to possible degradation of energy threshold, collection area and angular resolution.  Typical observations by IACTs require long integration times, hence the achievable effective duty cycle of the observatory (now energy and angular resolution dependent) becomes a key parameter for site selection. 
Very high clouds may lie entirely above the observed air showers and do not hinder observations above a certain threshold energy. 
Air turbulence or seeing, in turn, are not so critical since the requirements on the quality of the optics of IACTs are not comparable to those for modern optical telescopes. 
Finally, the prohibitive cost of providing domes for these large telescopes means that extreme weather phenomena such as low temperatures, high winds or sandstorms become crucial parameters for their design.
%Extreme weather phenomena like low temperatures, strong wind or sand storms become crucial parameters for the design of free-standing telescopes, not protected by a dome 
%which would come along with a prohibitive cost.
	
To help finding the best site for the CTA, all information about observation and climate conditions that is available from existing instrumentation 
including satellite data can be evaluated. However, satellite instruments do not always provide comparable measurements and require validation from the ground. 
Moreover, satellite data archives offer many possibilities, but  
comparing the candidate sites based on such datasets is not possible in some cases,
due to limited temporal and/or spatial resolution and the impact of site topologies, which can be difficult to interpret.

It therefore makes sense to cross-calibrate each candidate site using identical hardware, and 
for this purpose the CTA community decided to design and set up standardized sensor stations for collecting LoNS, cloud and weather data. 
With the remote location of some candidate sites, implying a lack of infrastructure and availability of experts for maintenance, 
the requirements on such an instrumentation are challenging: relative compactness, independence from power grid, easy maintenance and remote internet control. 
The devices should be combined into a compact and rugged station, providing off-grid power supply, computing power and possibly also a communication link to the Internet
or temporary storage and regular pick up of the data. 
Such requirements imply reliability resulting from mechanical simplicity, low power consumption and limited bandwidth. 
%This on the other hand limits the choice of instruments and sensors that can be used in such an environment.
The eventual choice of instruments was: a multi-purpose weather sensor, an infrared cloud sensor and specially developed sensors for measuring the LoNS. 
The basis and main instruments of this sensor station were developed and designed at the Max-Planck-Institute for physics (MPI) in Munich, Germany,
and later improved by the Centre de Physique des Particules de Marseille, France. 
The device was given the name Autonomous Tool for Measuring Observatory Site COnditions PrEcisely (ATMOSCOPE).
At the same time, the Joint Laboratory of Optics of the Palacky University Olomouc, Czech Republic, 
and the Institute of Physics of the Czech Academy of Science developed All-Sky-Cameras which were later incorporated in all but one ATMOSCOPE station.

%The LoNS sensor consists of a large area PIN diode and a filter wheel behind a lens, to make it a focal instrument with a defined angle of acceptance. 
%This device was designed for performing measurements in two wavelength bands, the first using a standard Johnson V-filter, mainly for calibration purposes, and 
%a second one using a combination of filters which matches best the spectral acceptance of the photomultipliers, foreseen for the CTA cameras, 
%covering most of the Johnson/Bessell B-filter and parts of the U-filter.

	%\section{General requirements on the ATMOSCOPEs}
	%
	%For an as low as possible Energy threshold, especially for the large size telescopes (LSTs), a very low level of light of the night sky (LoNS) is required. Another very important issue is the fraction of nights that are affected by cloud coverage. Apart from that, more general weather parameters like temperature, humidity, wind direction and speed are also interesting for us. Especially the wind speed is quite important, due to the large optical dishes that will be placed outside without a protective dome. Strong wind does not only endanger the structure, but can also make observations impossible, if the pointing of the telescopes is affected.\\
	%After all, the design for the ATMOSCOPE does include: General purpose weather sensor, IR cloud sensor and a special device for measuring the LoNS level. The other key role in the design is taken by the requirements on autonomous operation and independence of grid power supply.

	\section{Mechanical design and power supply}

\begin{figure}
	\begin{center}
		\includegraphics[width=0.48\textwidth]{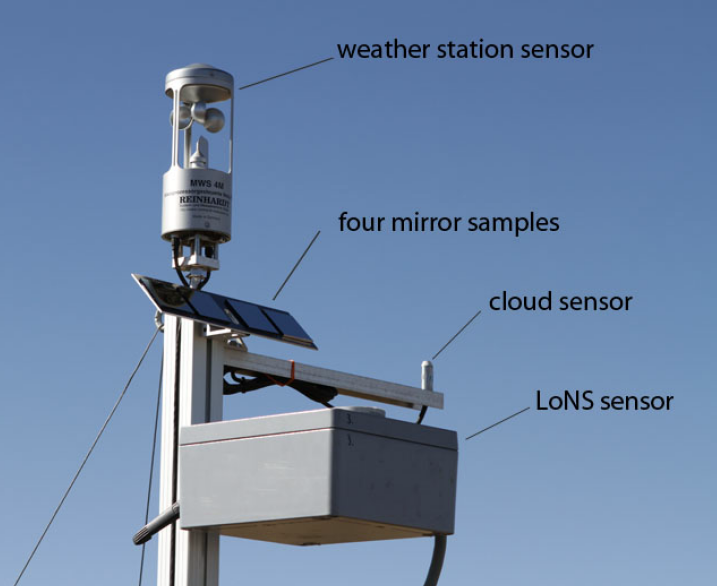}   
	\end{center}
	\caption{ATMOSCOPE instrumentation on top of the mast.}
	\label{fig:instr}
\end{figure}
\begin{figure}
	\begin{center}
		\includegraphics[width=0.48\textwidth]{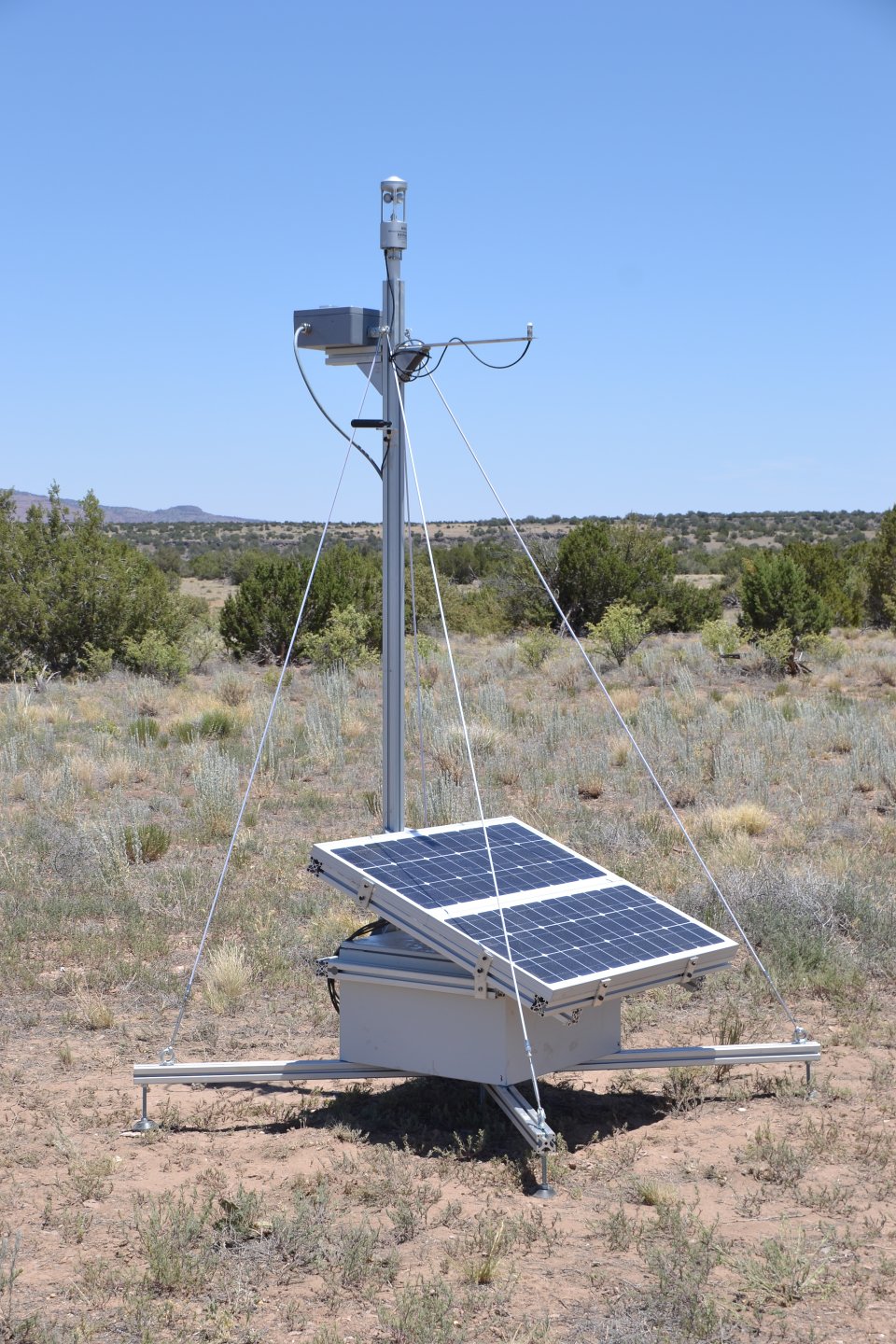}
	\end{center}
	\caption{%
Picture of an Autonomous Tool for Measuring Site COnditions PrEcisely (ATMOSCOPE), a long term site evaluation measurement station for CTA.}
	\label{fig:atmoscope}
\end{figure}
The primary requirement of the design of the site monitoring device was to construct a compact and robust station that can operate independently of the grid and under harsh outdoor conditions. 
%It was therefore decided in favor of a concept where 
The power is therefore provided by solar panels and stored in lead-acid batteries. 
%The batteries themselves are mounted, together with a charger and a control computer, inside a water proof electric control cabinet. 
The different instruments, shown in Figure~\ref{fig:instr}, are mounted on a mast $\np[m]{2.5}$ above the ground. 
The weather station and the cloud sensor were relocated to a separate 10~m mast for the final installation. 
Figure~\ref{fig:atmoscope} shows a picture of the installation, before relocation of the weather station and cloud sensor.
% assembled for the first time at MPI in Munich.

All the structural elements such as the mast and quadropod are made from $5$~cm~$\times$~5~cm profile aluminium tubes. The power for the ATMOSCOPEs is generated by solar panels providing a peak power of $\np[W]{100}$, which is used to charge two lead-acid batteries with a capacity of $\np[Ah]{100}$ each. All electric and electronic parts are housed in a cabinet suitable for outdoor electric installations, placed below the solar panels for additional protection from rain and 
%high temperatures during day-time due to
direct sunlight. The power system is designed to avoid shutdown even in the case of several days of bad weather: out of all the ATMOSCOPEs operated over several years, the batteries 
were exhausted on one site only, leading to a cumulative loss of 26~hours. The cause was persistent bad weather in winter.
%were exhausted on one site only, during a cumulative period of 26~hours. The cause was persistent bad weather in winter. 

The mast is kept in place by four steel ropes under tension, going to the tips of the quadropod. 
The whole station is to be set up in a leveled position with the mast side pointing towards the equator. The orientation of each ATMOSCOPE with respect to the local coordinates is hence fixed, and can later be used for the analysis of star light captured by the LoNS sensors.
% ANSWER TO COMMENT (by Jean-Pierre) 
The precision of the alignment of the optical devices is of the order of one to a few degrees, depending on the site. This characteristic can be measured by comparing the evolution of the light of bright stars passing in the field of view with the expected one.
	
\section{Weather station with cloud sensor}

\begin{table}\footnotesize
\centering
\begin{tabular}{ c | c }
\hline
measured parameter & technical properties \\
\hline \hline
%	weather station & 
%	\begin{tabular}{ c c }
	  \begin{tabular}{ c } temperature \\ pressure \\ humidity \\ wind speed \\ wind direction  \end{tabular} &
	  \begin{tabular}{ c } -40 to $\np[\deg C]{60}$ ($\pm \np[\deg C]{0.3}$) \\ 600 to 1000~hPa ($\pm 0.8$~hPa) \\ 1 to $\np[\%]{100}$ ($\pm \np[\%]{2}$) \\ 0 to $\np[km/h]{150}$ ($\pm \np[km/h]{2}$), starting at 0.5~m/s \\ 0 to 360$^\circ$ ($\pm 5^\circ$, hysteresis $<8^\circ$, starting at 0.5~m/s)\end{tabular} \\
%	  \end{tabular} \\
	  \hline
%  	  cloud altitude at night-time & 
%	  \begin{tabular}{ c c }
%	    \begin{tabular}{ c } sensitivity after cross-calibration \\ sensitivity after correction for temp. and RH \\ cloud altitude resolution \end{tabular}
%	    \begin{tabular}{ c } $0 .. \np[km~a.g.l.]{4}$ \\ $4.5-7.3$~km~a.s.l.  \\ $\pm \np[km]{0.3}$ (above 1~km)\end{tabular}
%	  \end{tabular} \\
%	  \hline
\end{tabular}
\caption{Summary of different instruments and the most important technical details of the weather station.}
\label{tab:instr_sum}
\end{table}

For collecting the basic weather parameters relevant for site selection, a compact, commercially available weather station was selected. The decision was made to employ a Reinhardt MWS4~\citep{mws4}, which can be read out via a RS232 interface and measures temperature, barometric air pressure, relative humidity, wind speed and direction (see table~\ref{tab:instr_sum}). 
In addition, it was equipped with a cloud sensor offered by the same company~\citep{cloud},
which uses a thermopile to measure the difference between the ambient temperature and the radiation temperature of the sky. 
It then infers the altitude of a cloud assuming a constant temperature lapse rate of $-6.5$~K/km.

The interval at which weather  data were taken, initially set at one minute as for the LoNS
measurement, was decreased to 2~seconds after a few months of operation in order to better estimate the time profile of weather parameters, especially the wind.
%A value estimating the cloud base is already computed by the weather station, however needs to be calibrated additionally on each site, 
%for dependencies on ambient temperature and relative humidity. 
%Humidities are reliably measured in a range from 2\% to 100\%, with 2\% accuracy, while the wind speeds 
%are measured from about 2 to 150~km/h, with $\pm$2~km/h measuring accuracy. The wind direction is measured, within this range of wind speeds, with $\np[\deg]{5}$ accuracy 
%and $<\np[\deg]{8}$~hysteresis.

IR radiometer measurements, such as that from the thermopile, are affected by changing temperature profiles, dry or wet adiabatic lapse rates and cloud or haze layers, which may be too thin and not 100\% opaque. 
The cloud sensor has a full field of view of the order of $\np[\deg]{10}$, thus for an inhomogeneous cloud coverage, only a mean value is measured. 
Saturation is reached for an estimated cloud altitude of $\sim\np[km]{7}$; however, experience with the installed instruments showed that only clouds below about 4000~m above ground could be detected 
reliably, depending on the site.

\begin{figure}[h!]
	\begin{center}
		\includegraphics[width=.49\columnwidth]{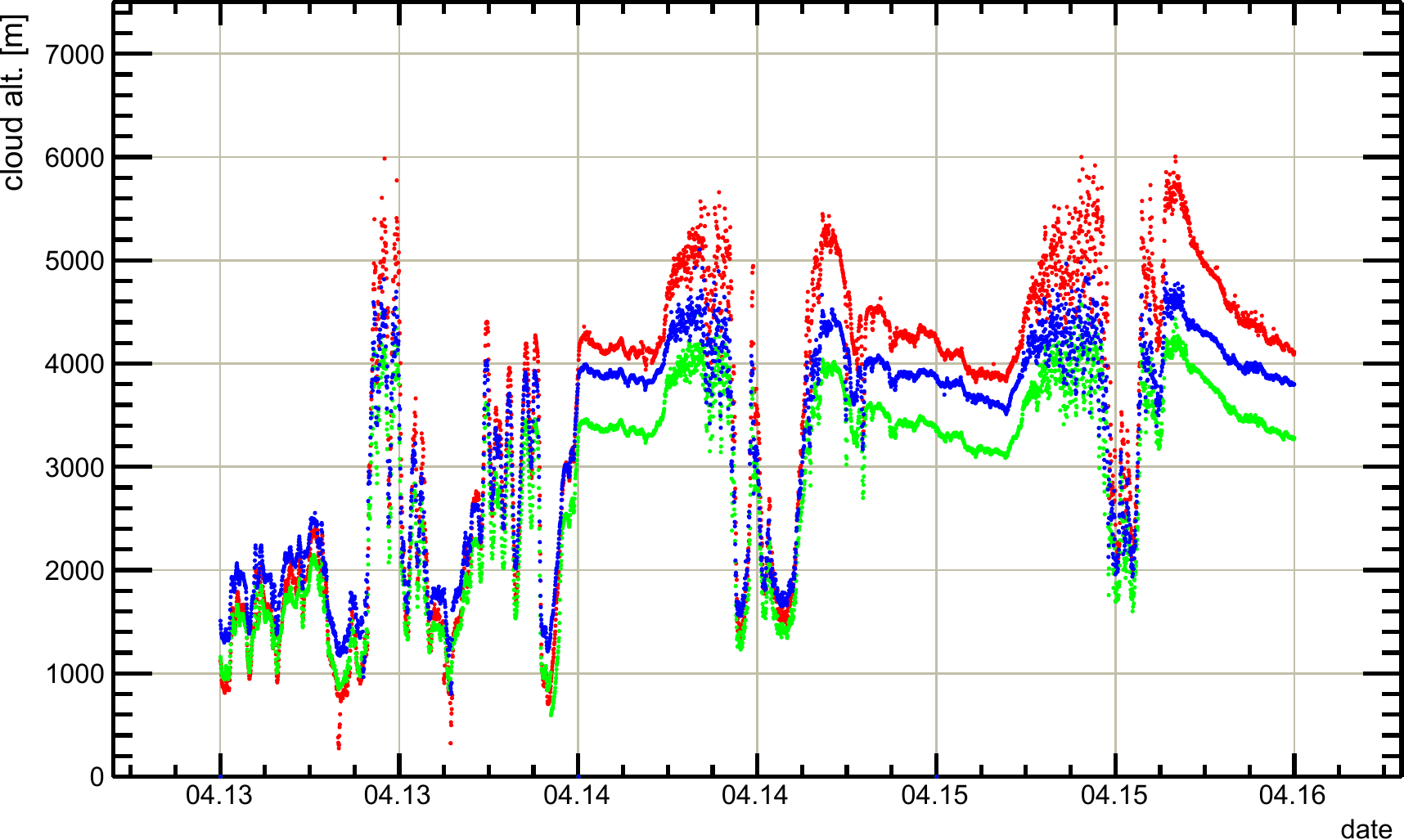}
		\includegraphics[width=.49\columnwidth]{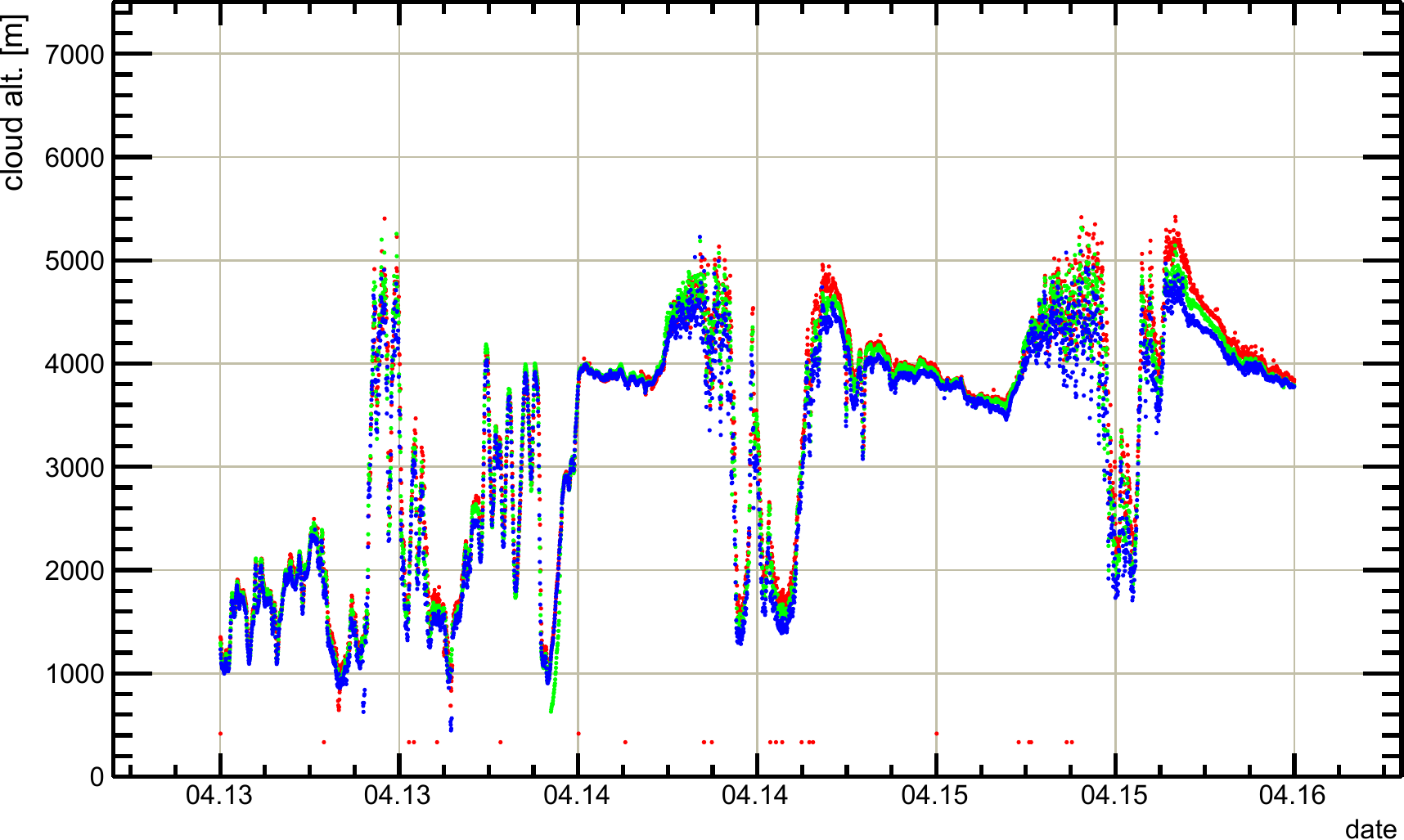}\\
		\includegraphics[width=.49\columnwidth]{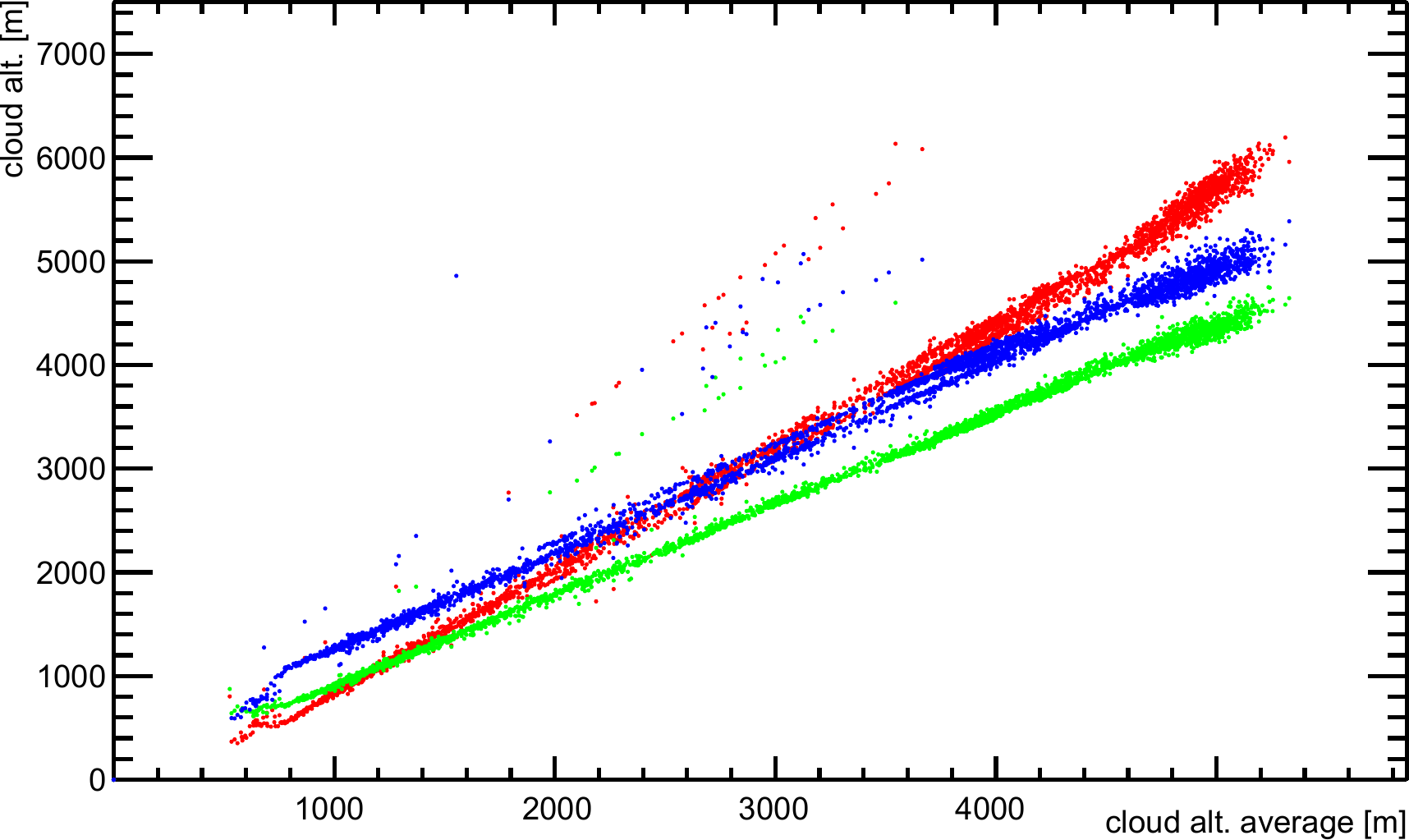}
		\includegraphics[width=.49\columnwidth]{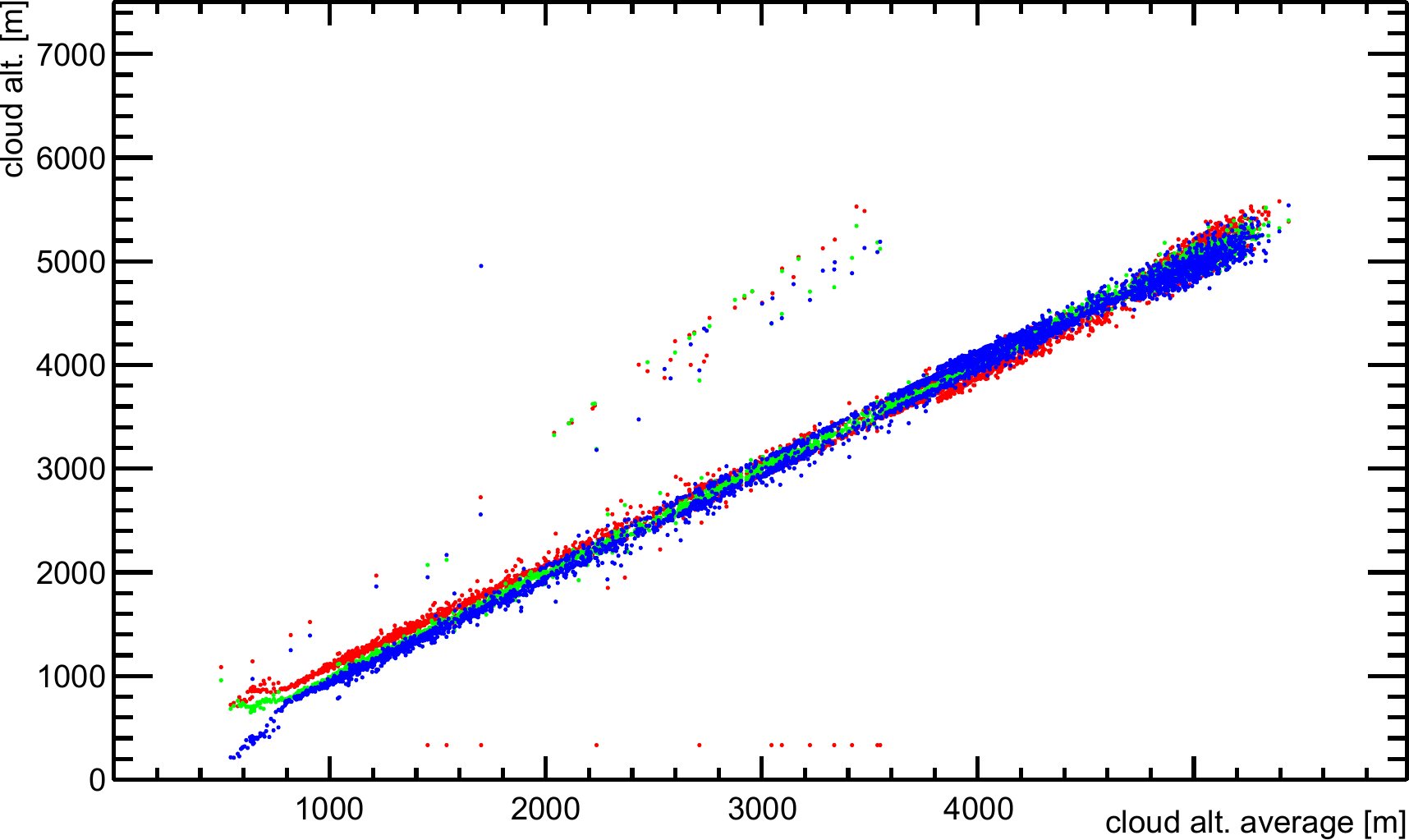}
	\end{center}
	\caption{%
		Thermopile cloud altitude measurements vs time (top) and mean of the 3 sensor measurements (bottom), before (left) and after (right) cross-calibration. The individual devices show considerable measurement differences, but these follow nicely a linear correlation allowing corrections to be made.}
	\label{fig:cloud_corr}
\end{figure}

The cloud sensors have been cross-calibrated against each other by placing them next to each other and exposing them to the same sky. The first such cross-check between 3 ATMOSCOPEs at MPI in Munich served as the basis for the overall cross-calibration. Figure~\ref{fig:cloud_corr} shows the data of about 3 days of measurements during mostly covered skies from the cross-check dataset before and after correction for the different response of the individual sensors. By plotting the measured cloud altitude of each instrument against the mean of the 3 sensor measurements a linear relationship became apparent. The data from each cloud sensor were therefore treated with a linear function, such that all reproduce the same mean value. This procedure yields only a relative calibration; on the other hand we were not too interested in precision with respect to the cloud height, but rather were aiming to measure the average cloud coverage.

\begin{figure}
\centering
\includegraphics[width=0.49\columnwidth]{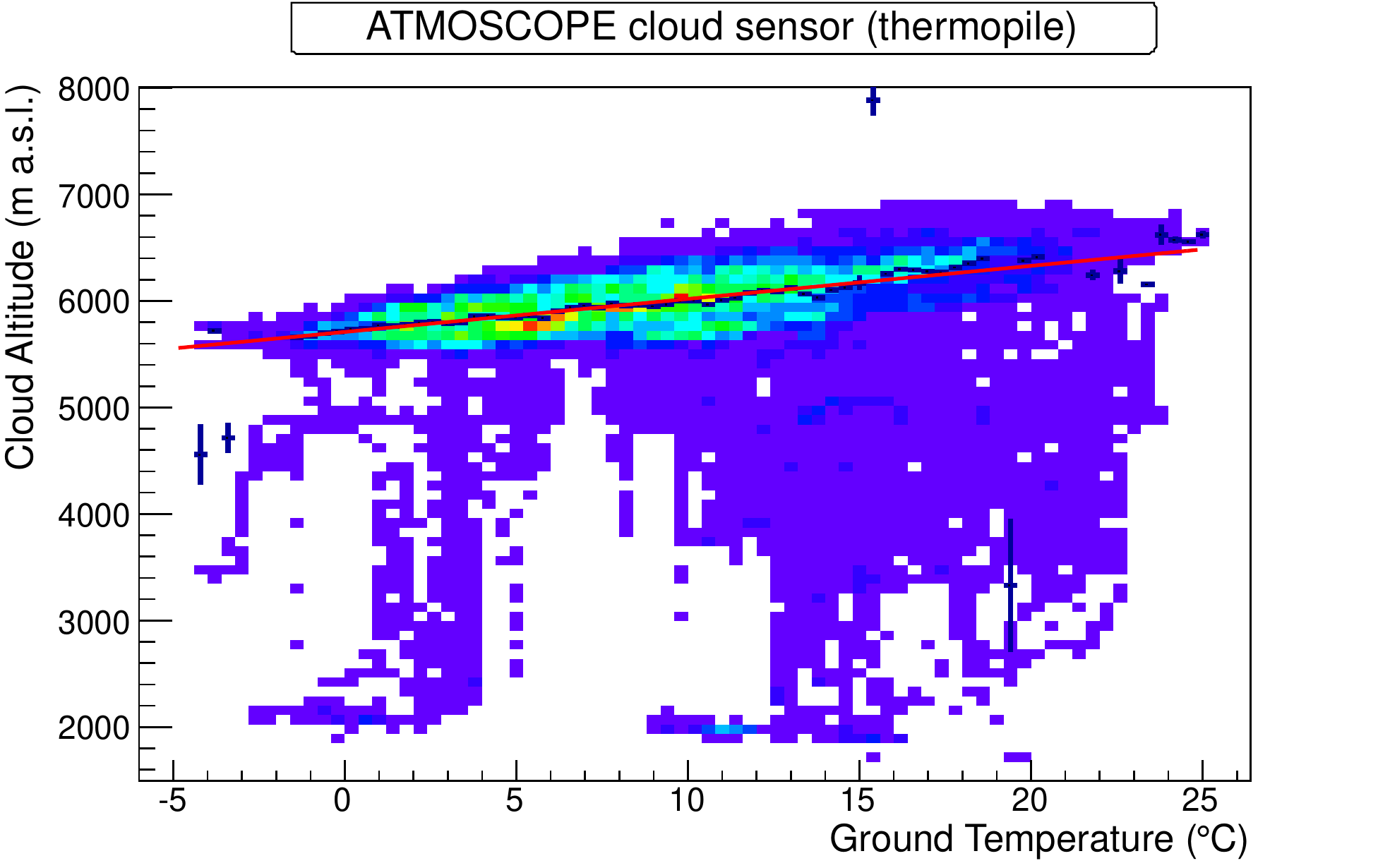}
\includegraphics[width=0.49\columnwidth]{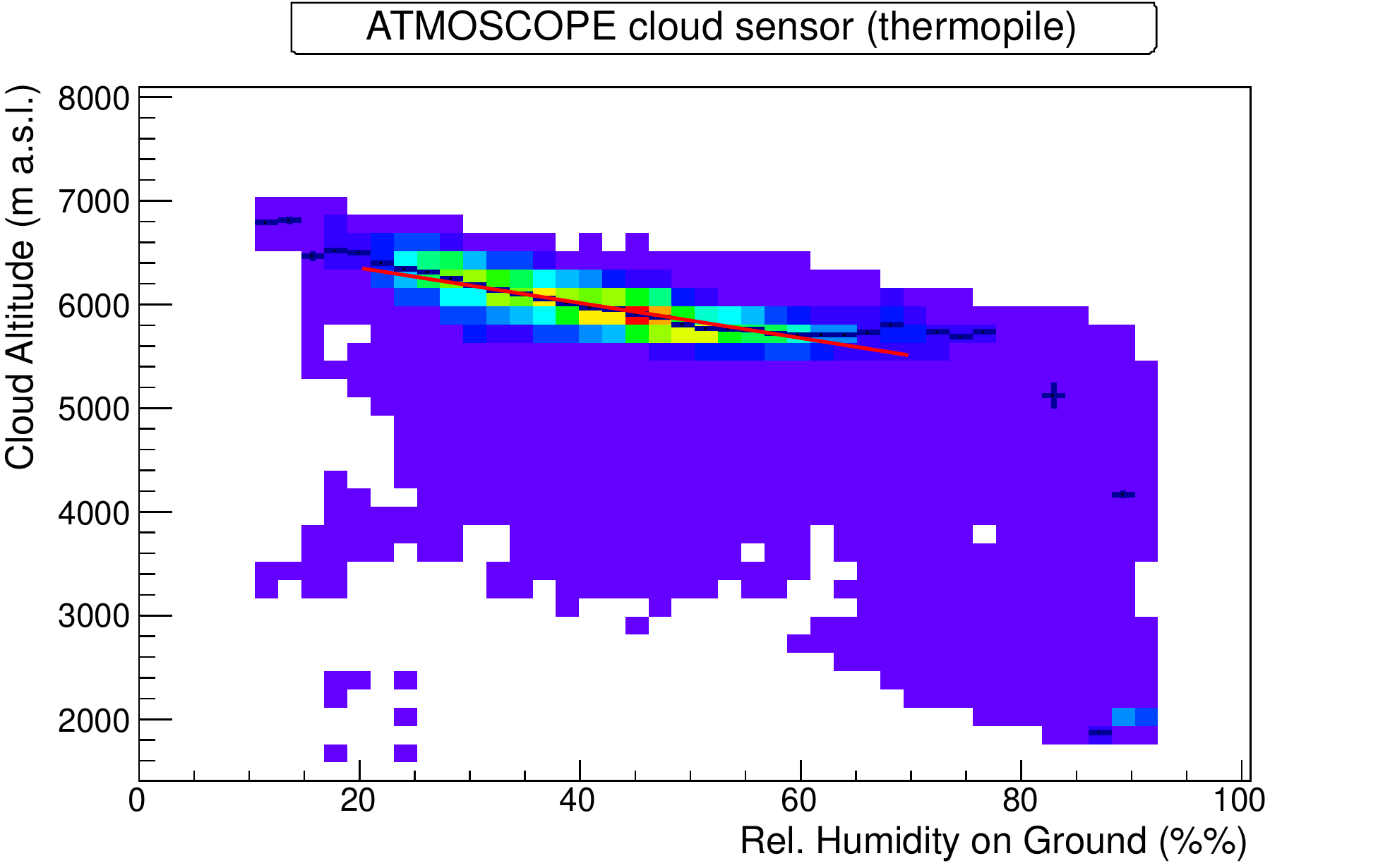}
\includegraphics[width=0.49\columnwidth]{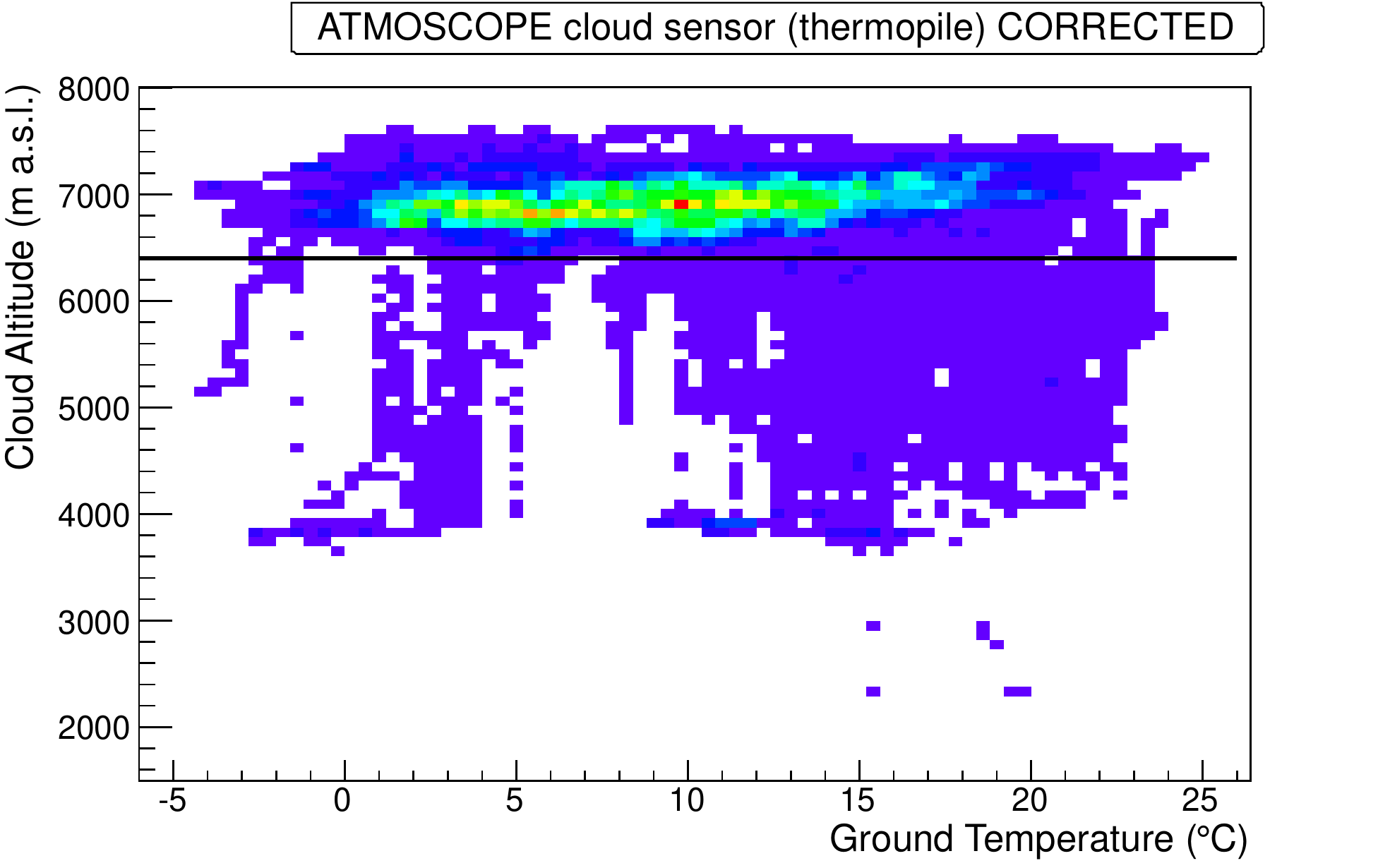}
\includegraphics[width=0.49\columnwidth]{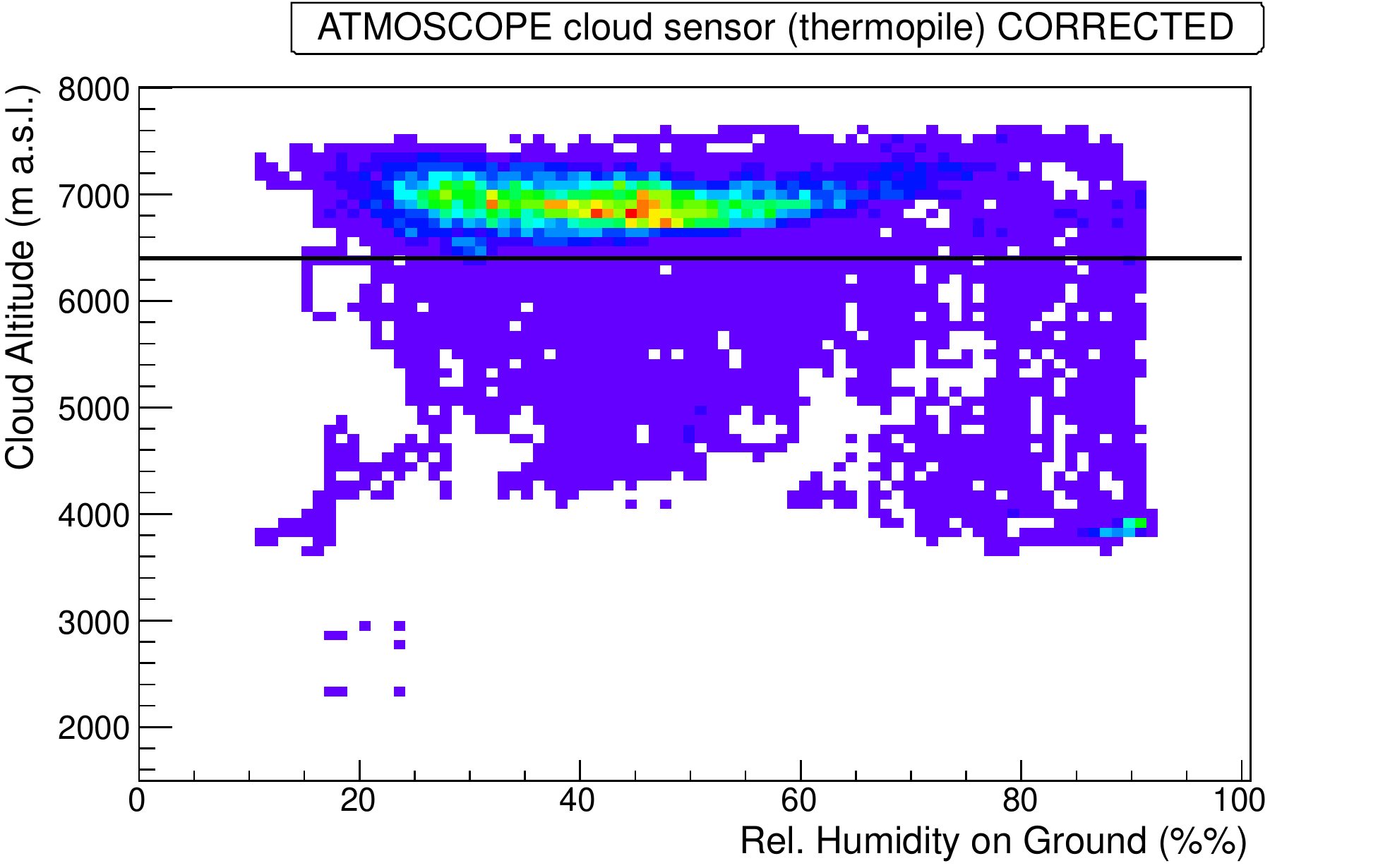}
\caption{Top: Cloud heights, obtained from one deployed Reinhardt thermopile during a half-year measurement campaign, 
as a function of ground temperature (left) and relative humidity on the ground (right). 
Bottom: The corrected cloud heights, plotted against temperature and relative humidity. The coloured entries correspond overwhelmingly to cloudless nights, hence 
the black line shows the assumed sensitivity limit of this device.}
\label{fig:clouds}
\end{figure}

\begin{figure}
\centering
\includegraphics[width=0.49\columnwidth]{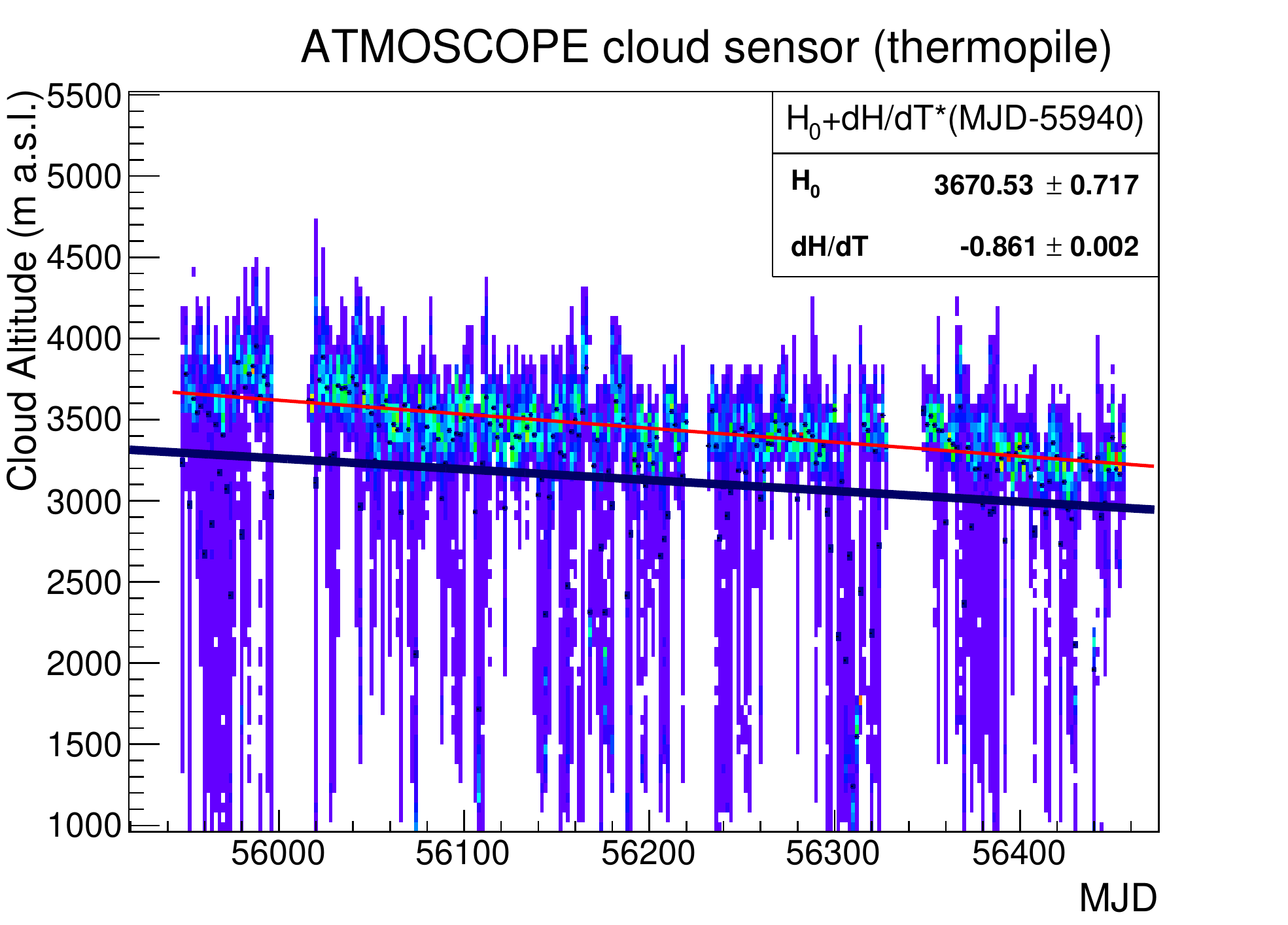}
\includegraphics[width=0.49\columnwidth]{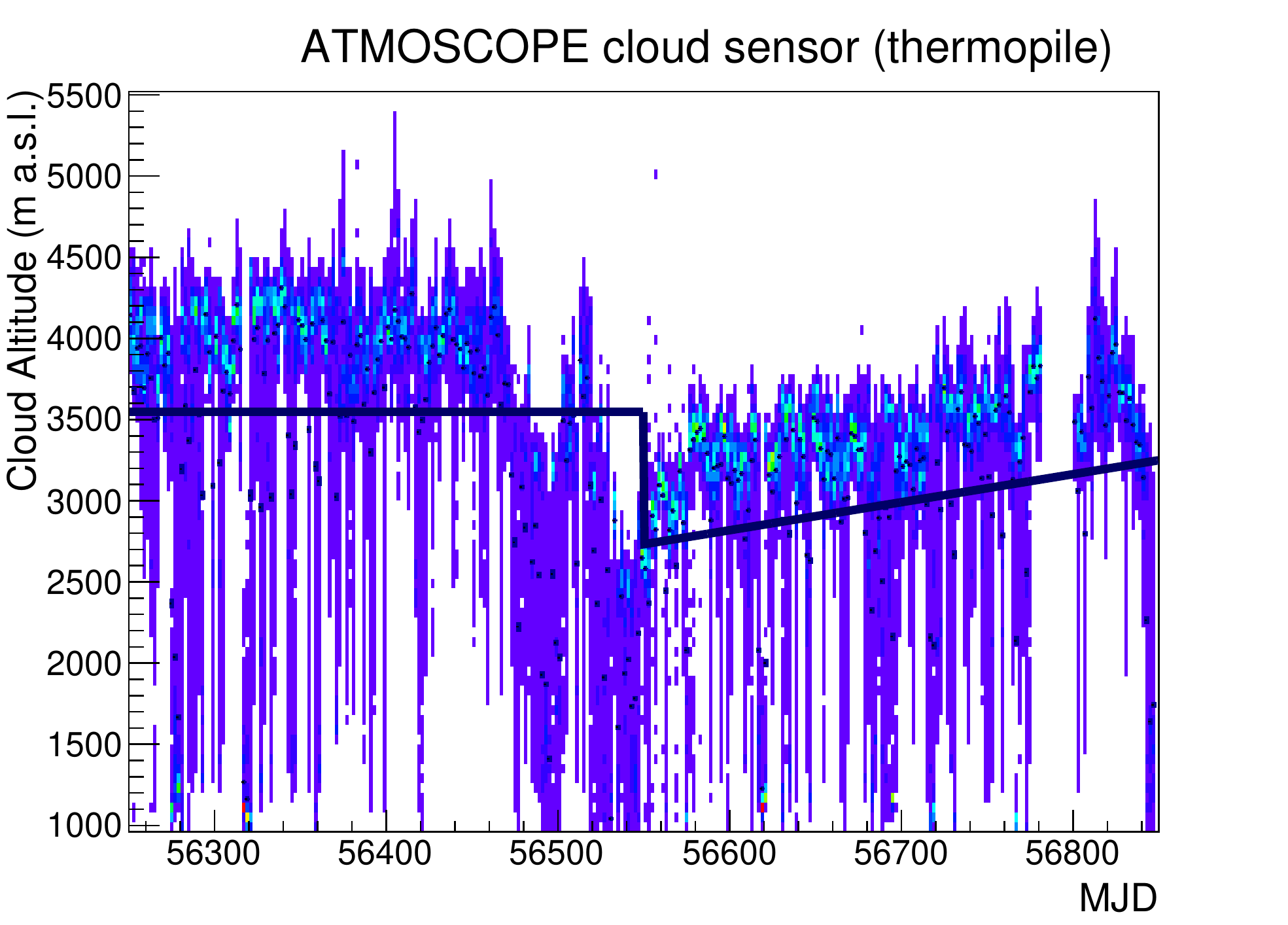}
\caption{Cloud heights, obtained from two deployed Reinhardt thermopiles as a function of time, expressed as Modified Julian Days (MJD). 
The coloured entries show the measured cloud height and the black line the derived sensitivity limits of the devices. 
While on the left side a gradual decrease of sensitivity with time is observed, mainly attributed to dust, the right side shows a two months period of frequent 
rain, starting around MJD$=56460$. At the end of that period, the thermopile has lost sensitivity which it slowly recovers during the next 250 days. That the drop of 
sensitivity was not mimicked by clouds has been cross-checked with images from the All-Sky-Camera.}
\label{fig:cloudstime}
\end{figure}

Nevertheless, once installed at the sites, we discovered that residual dependencies of the cloud height on at least ground temperature and relative humidity exist, which need to be corrected for as well. Figure~\ref{fig:clouds} (top) shows an example of such a dependency. The retrieved cloud height maxima (corresponding to cloudless nights) were fitted with a linear function with respect to both parameters and corrected (see Figure~\ref{fig:clouds} bottom). The temperature gradients obtained  range from -25~m/K to +65~m/K for the different ATMOSCOPEs, while the relative humidity gradients are always negative, ranging from -25~m/\% to -4~m/\%. After correction, each ATMOSCOPE was found to be sensitive to quite different maximum cloud heights, ranging from 4500 to 7300~m a.s.l. 

Tests in the lab had revealed that the measured cloud altitude depends both on water and dust deposits on the lens. The thermopiles were never cleaned on-site, and typically 
show sensitivity degradations of roughly 1~m per day with time (see Figure~\ref{fig:cloudstime} left). 
%After long periods of rainfall, water deposits may form on the lens which evaporate only slowly and cause temporal degradations of sensitivity. 
During long periods of rainfall, wet dust deposits may solidify on the lens and then disappear only slowly and cause temporary degradation of sensitivity.
Figure~\ref{fig:cloudstime} (right) shows such an example.

While this method offers data for relatively low clouds at low cost and low instrumental complexity for comparative studies of the average cloud coverage during the nights, 
the obtained wide range of sensitivities caused some concern about the feasibility of such studies. Consequently, the device was finally only used for rough estimates. 

\section{All-Sky-Camera}

\begin{figure}[h!]
	\begin{center}
		\includegraphics[width=0.5\linewidth]{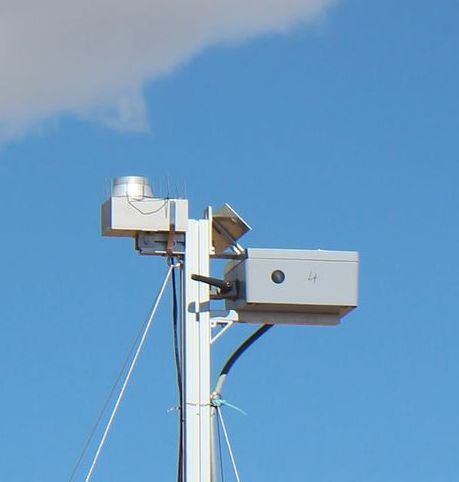}
	\end{center}
	\caption{Picture of an ASC (left) deployed opposite the LoNS sensor (right). The ASC is protected against birds by wires sticking out of the box. }
	\label{fig:ascspm}
\end{figure}

A more precise measurement of the cloud coverage during nights is achieved with the use of All-Sky-Cameras (ASCs), installed at each candidate site on the ATMOSCOPE structure,
opposite the LoNS sensor\footnote{This is true except for one site where an ASC of a different type was already installed.} (see Figure~\ref{fig:ascspm}).
%Its design and software were developed by the 
%Eight ASCs have been installed and measure night sky parameters at eight of the candidate sites (except HESS) of the Cherenkov Telescope Array (CTA) gamma-ray observatory.
%The HESS site is not excluded from analysis the images of the HESS ATOM camera was analysed using the same algorithm as ASC.
The ACS system consists of a fish eye lens of type FUJINON YV2.2x1.4A-SA2~\citep{fujinon},
an astronomical CCD camera of type G1-2000~\citep{ccd},
 a control computer and associated electronics. The ACS is fed by the power supply of the ATMOSCOPE and uses the same internet 
connection, but the measurements are taken independently and do not interrupt measurements by the other devices of the ATMOSCOPE.
The CCD is equipped with a ICX274AL chip  which has a quantum efficiency greater than 50\% in the range from 450-550~nm. 
Its resolution is $1600 \times 1200$ pixels, and the pixel depth is 16-bit monochrome.  
The fish-eye varifocal lens has a 185$^\circ$ field of view, and its iris can be controlled electronically.   
It is covered by a ``zero'' meniscus lens, correcting for the divergence of the incident beam. 

The camera and lens are housed in a solid aluminium waterproof body and connected to a miniPC via USB and a separate serial I/O controller, 
which controls the iris and the power switch of the camera. 
The optical switching electronics turns the system ON after sunset and keeps it going during astronomical night-time. 
The system is OFF during the day, in order to save energy. 

\begin{figure}[h!]
	\begin{center}
		\includegraphics[width=0.99\linewidth]{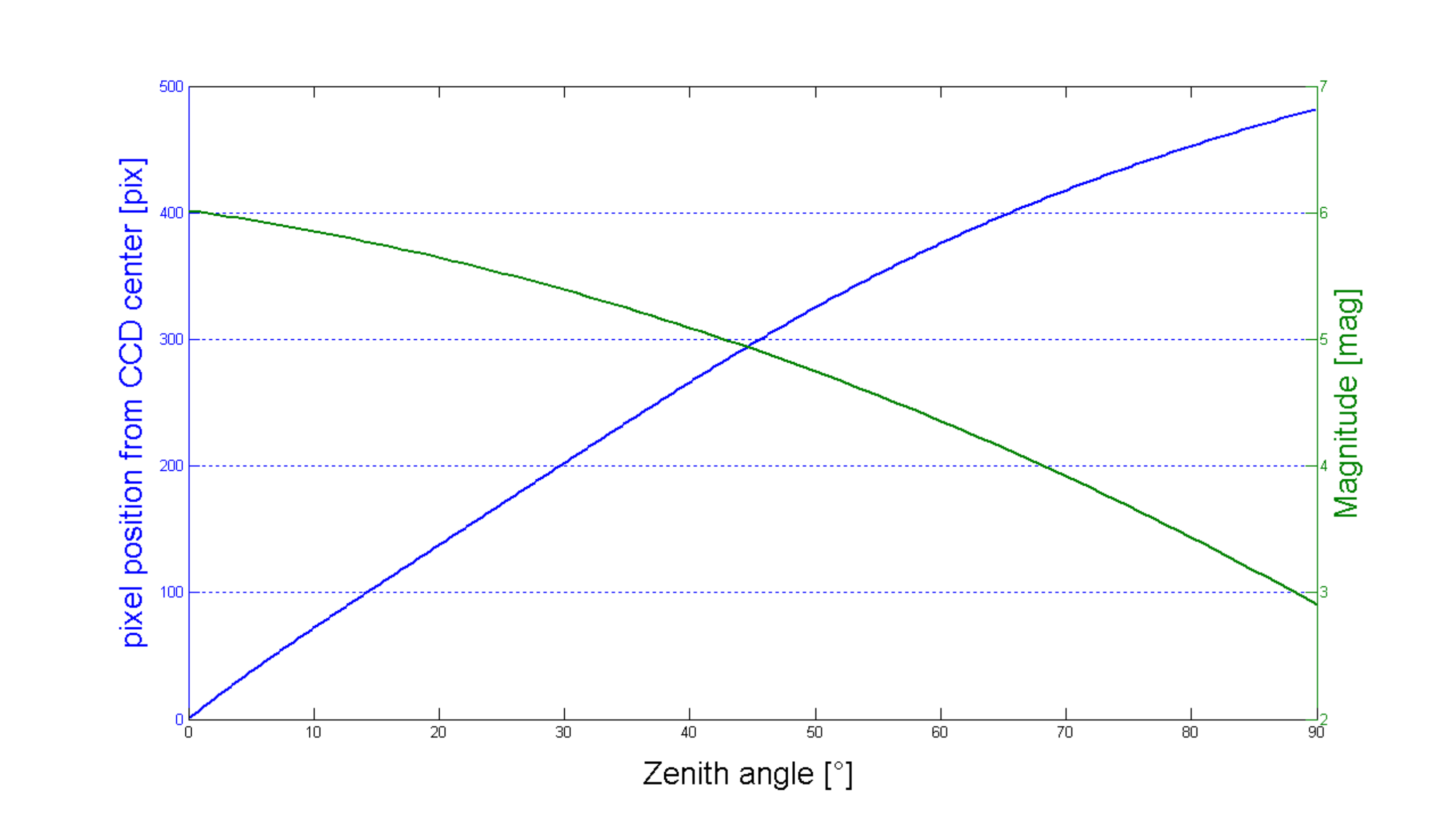}
	\end{center}
	\caption{CCD positon and star magnitude limit vs. zenith angle, as obtained from the ASC calibration.}
	\label{fig:asccalib}
\end{figure}

Images are taken automatically every five minutes and processed using the control computer of the device. 
The resulting analysis output is the cloud fraction (CF), defined as the percentage of the sky covered by clouds. 
The data are analysed on-site, and only the results are sent to the data server using the ATMOSCOPE internet connection.

%The power system is independent and usually supplied with one 250pW panel solar charger and 100A battery. 
%The ASCs use the same Atmoscope connection to the internet.

%These devices are described in detail elsewhere~\citep{asc2013}.

A calibration of the system's fish-eye and zero meniscus lenses was performed in the lab to correct for optical aberrations and vignetting effects. 
For this aim, a point-like optical source (a spot as small as possible) was placed at a distance of at least 10~m from the lens. 
Then, the body of the camera was rotated around the CCD detector's sagittal axis from 0 to 180 degrees. 
This procedure was repeated for the transverse axis. Intensity and pixel position of the light source as a function of incidence angle were then calculated 
(see Figure~\ref{fig:asccalib}). 

\begin{figure}[h!]
	\begin{center}
		\includegraphics[width=0.99\linewidth]{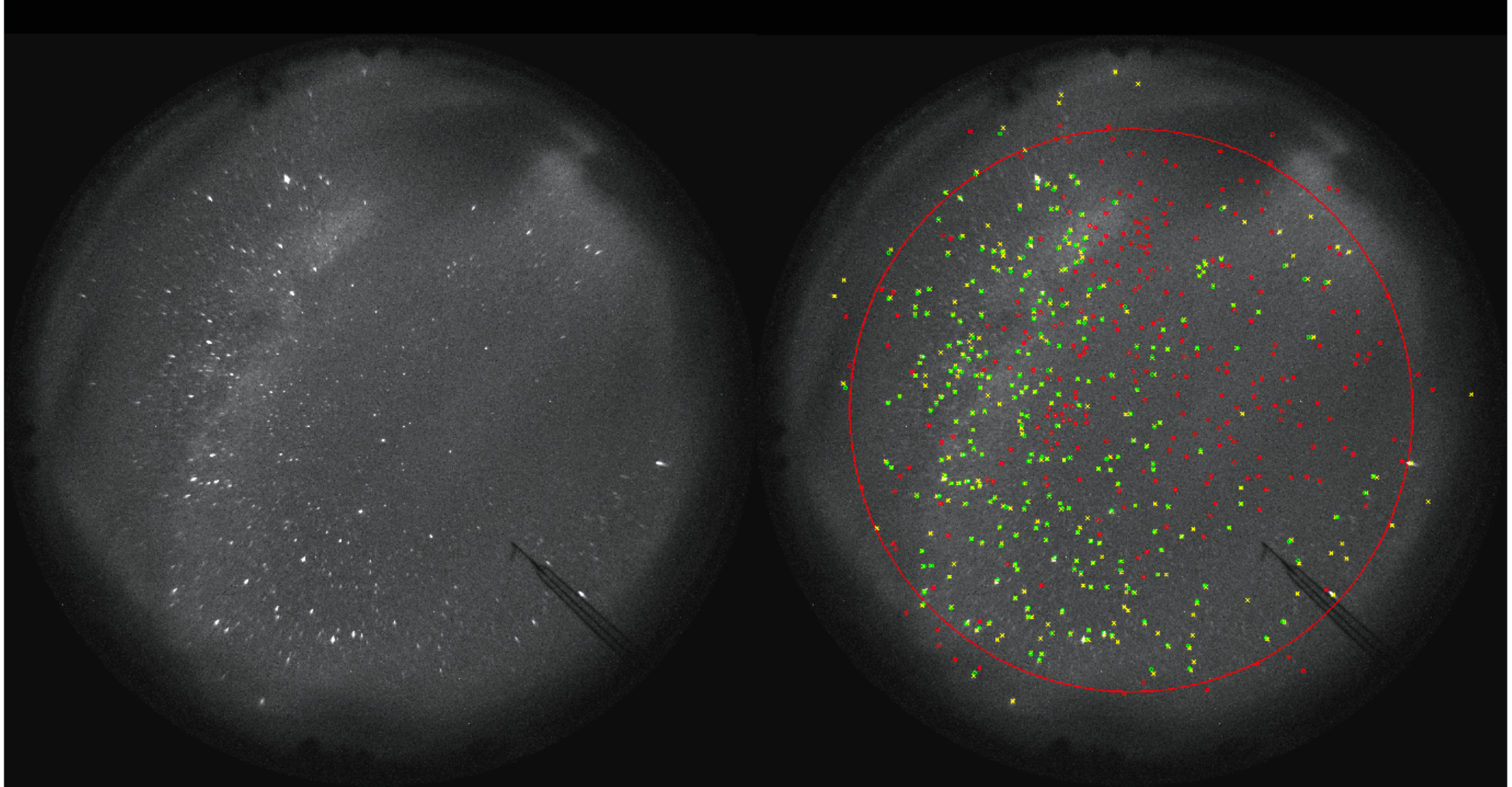}
	\end{center}
	\caption{An example of the cloud analysis of an ASC image of the sky partly covered by clouds. 
Left: raw image without analysis, right: with the result of the analysis superimposed.
The yellow crosses show all detected stars, out of which the green squares show those detected stars whose position matches a catalogue star; 
the small red circles depict catalogue stars without correspondence in the ASC image. These are typically obscured by clouds. 
In the lower right area of the images, the 10~m mast with the weather station is visible. The large red circle shows the analysis limit of 60$^\circ$ zenith angle.}
	\label{fig:asc}
\end{figure}

\begin{figure}[h!]
	\begin{center}
		\includegraphics[width=0.99\linewidth]{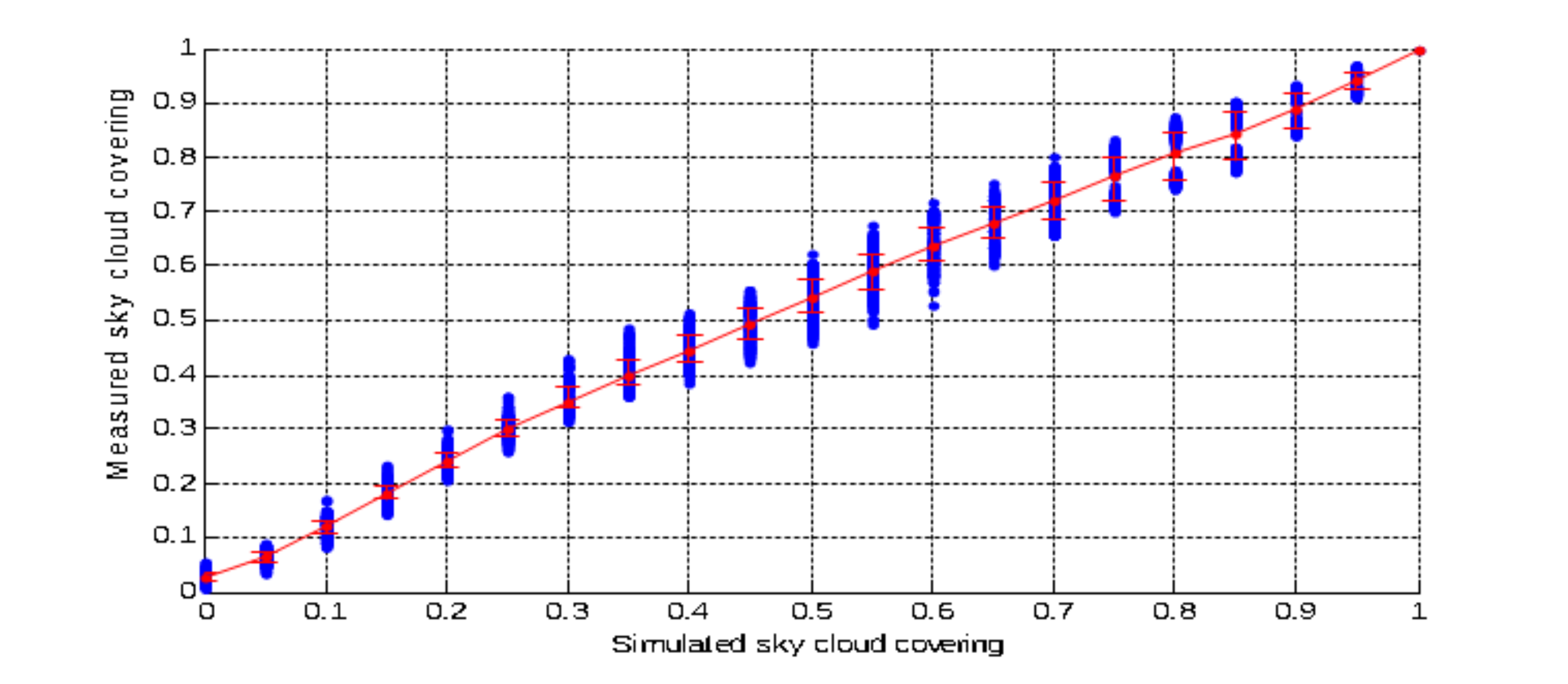}
	\end{center}
	\caption{Dependence of the cloud cover retrieved vs. simulated one. The blue points show all trials while the red points with error bars show the mean and RMS for each 
simulated CF bin. The red points have been linearly interpolated; the curve is used as a correction function to avoid small residual biases.}
	\label{fig:ascerror}
\end{figure}

The cloud fraction was defined as the ratio of the number of recognized image stars and catalogue stars in a field of view from 0 to 60$^\circ$ zenith angle. 
To estimate the second number, we used the visual magnitudes~\citep{bsc5}.
The catalogue stars have been filtered with respect to their magnitude and calibration. 
A magnitude limit for stars close to 60$^\circ$ zenith angle of slightly larger than  $4^\mathrm{m}$ was chosen relaxing gradually towards zenith, where $>6^\mathrm{m}$ was applied.
For each star in the filtered catalogue, we search for a detected star within an angular distance of $<1^\circ$ (see Figure~\ref{fig:asc}).
The ratio of paired divided by unpaired stars gives then the CF. 
Typically about 900 catalogue stars are tested over the whole observed night sky. 
The precision of the algorithm was estimated using artificial cloud simulations which yielded an uncertainty of the calculated CF of always less than 5\% in absolute CF, 
depending slightly on the simulated CF (see Figure~\ref{fig:ascerror}).
 
\section{LoNS monitor}

The main instrument of the ATMOSCOPEs for measuring the night sky brightness is the LoNS sensor. To avoid too much complexity and keep the data volume low, an integrating rather than imaging type of device was used. The final design uses imaging optics to map a certain part of the sky onto a large area PIN diode and measure the integral light flux via the photo current. A starlight model predicting the amount of ``natural light'' from this field of view can then be used to quantify the fraction of light pollution from artificial light sources as well as to detect clouds inside the field of view, which also modify the light flux.

\begin{figure}[h!]
	\begin{center}
		\includegraphics[height=0.38\textheight]{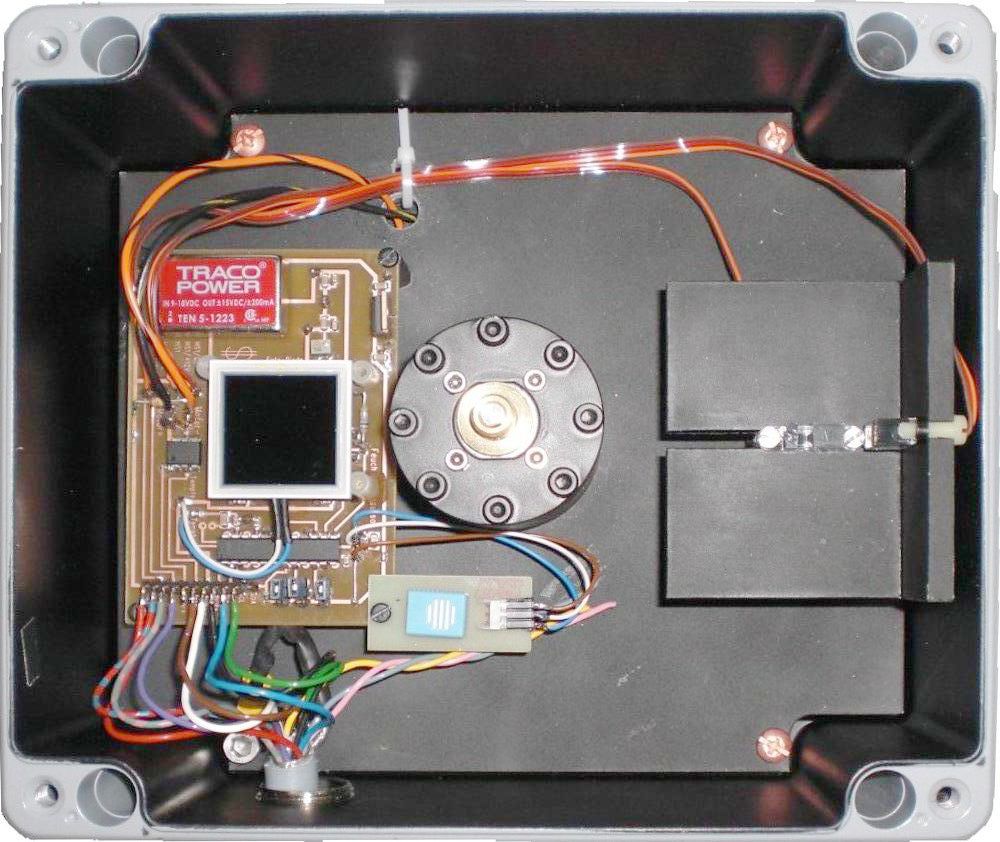}
	\end{center}
	\caption{LoNS measurement device, open box with the filter wheel removed. 
From left: PCB with PIN diode, motor for turning the filter wheel, Hall sensors for checking the position of the wheel.}
	\label{fig:lons_box}
\end{figure}

\begin{figure}[h!]
	\begin{center}
		\includegraphics[height=0.25\textheight]{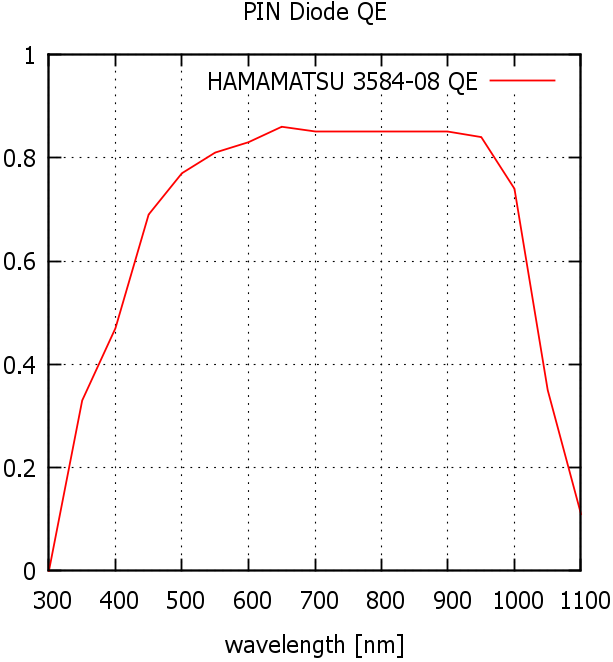}
    \includegraphics[height=0.25\textheight]{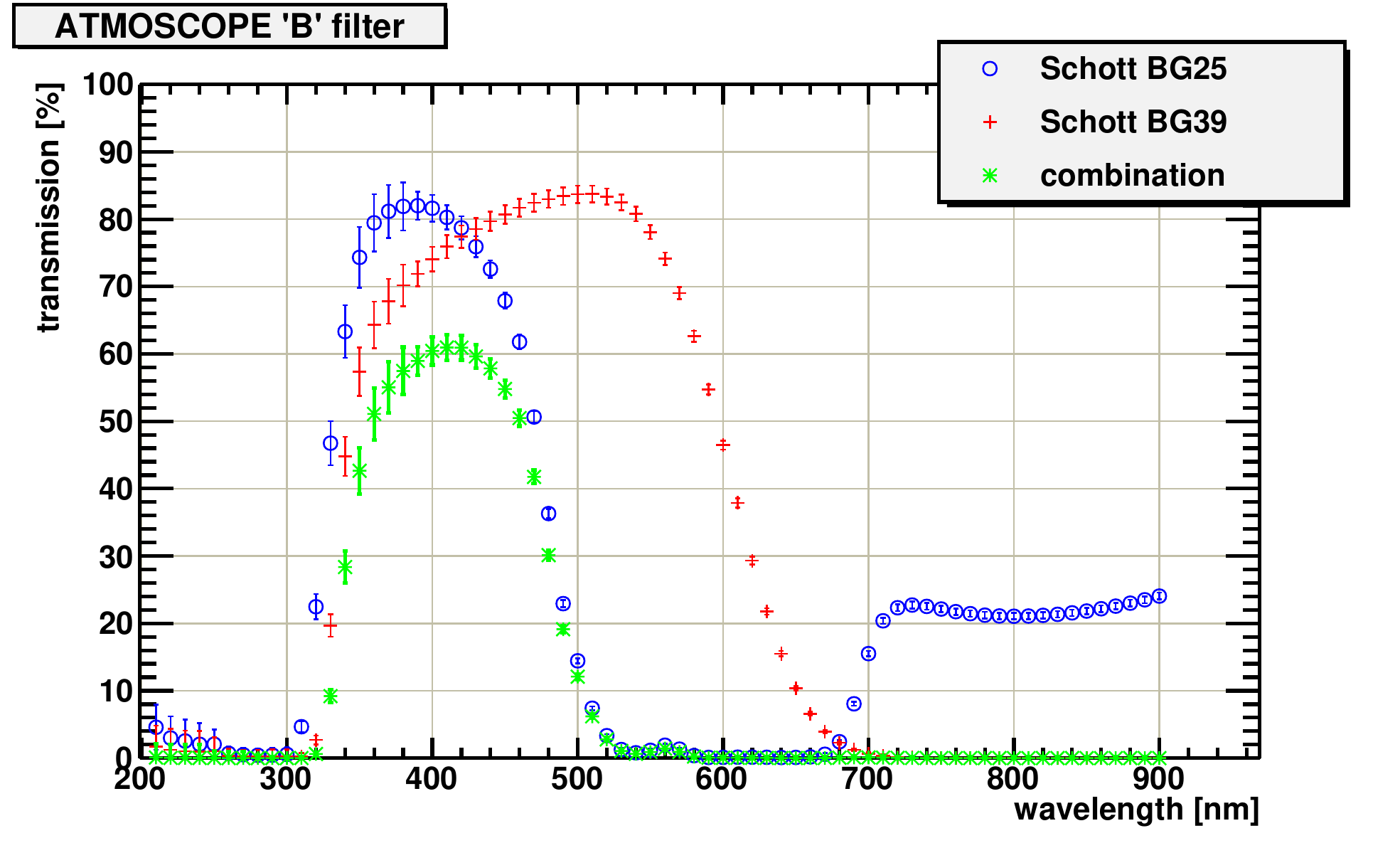}
	\end{center}
	\caption{Quantum efficiency of the Hamamatsu S3584-08 PIN diode, measured with the MPI Quantum Efficiency Measurement Device (left).
Transmission curves of the two filters, which have been combined to achieve the ``ATMOSCOPE-B'' filter for the LoNS sensor (right). }
	\label{fig:lons_pin}
\end{figure}

The light sensor is a $28 \times 28$~mm Silicon PIN diode of type S3584-08, 
manufactured by Hamamatsu~\citep{hamamatsu} (see Figure~\ref{fig:lons_box}). 
The quantum efficiency curve of the device, measured in the lab at MPI, is shown in Figure~\ref{fig:lons_pin}. The PIN diode is operated in photo-voltaic mode. 
The photo current is amplified by a trans-impedance amplifier with $\np[VA^{-1}]{E8}$ in the first stage and by two optional secondary stages 
with an amplification of 10 and 100 maximizing the dynamic range.
%For precision measurements of very weak current signals, as in our first amplification stage, the requirements on the amplifier are that it should have a very low input-bias current and a low temperature drift. 
The precise measurement of very weak currents in our first amplification stage requires an amplifier with a very low input-bias current and a low temperature drift.
A device that offers these properties is the OpAmp AD549~\citep{ad549} from Analog Devices. The manufacturer claims an input bias of $\np[fA]{250}$ and a temperature offset drift of $\np[\mu V K^{-1}]{5}$. A capacity of $\np[nF]{1}$ in the loop-back of the amplifier guarantees an integration time $RC = \np[s]{0.1}$, to avoid noise pickup and oscillations. 
%The light sensor is a $28 \times 28$ mm Silicon PIN diode, manufactured by Hamamatsu. The quantum efficiency curve of the device is shown in figure \ref{fig:lons_pin}. This PIN diode is operated in photo-voltaic mode. The photo current is amplified by the trans-impedance amplifier with $\np[VA^{-1}]{E8}$ in the first stage and by two optional secondary stages with an amplification of 10 and 100 maximizing the dynamic range.
%For precision measurements of very weak current signals, like in our first amplification stage, the requirements on the amplifier are: very low input-bias current and a low temperature drift. A device that offers these properties is the OpAmp AD549 from Analog Devices. The manufacturer claims an input bias of $\np[fA]{250}$ and a temperature offset drift of $\np[\mu V K^{-1}]{5}$ \cite{analog}. A capacity of $\np[nF]{1}$ in the loop-back of the amplifier guarantees an integration time $RC = \np[s]{0.1}$, to avoid noise pickup and oscillations. 

\begin{figure}[h!]
	\begin{center}
		\includegraphics[width=0.6\columnwidth]{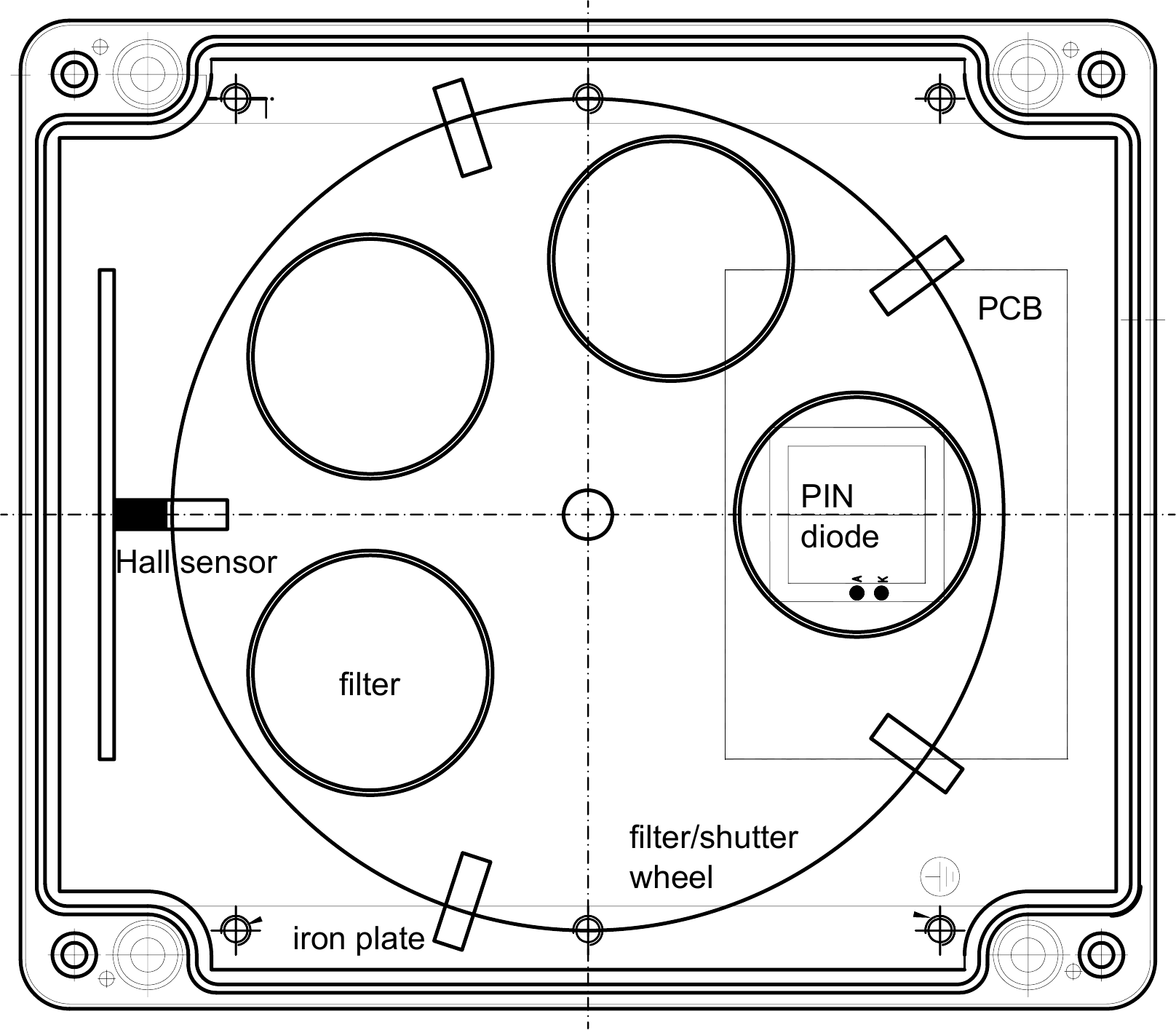}\\
		~\vspace{1cm}~ \\
		\includegraphics[width=0.6\columnwidth]{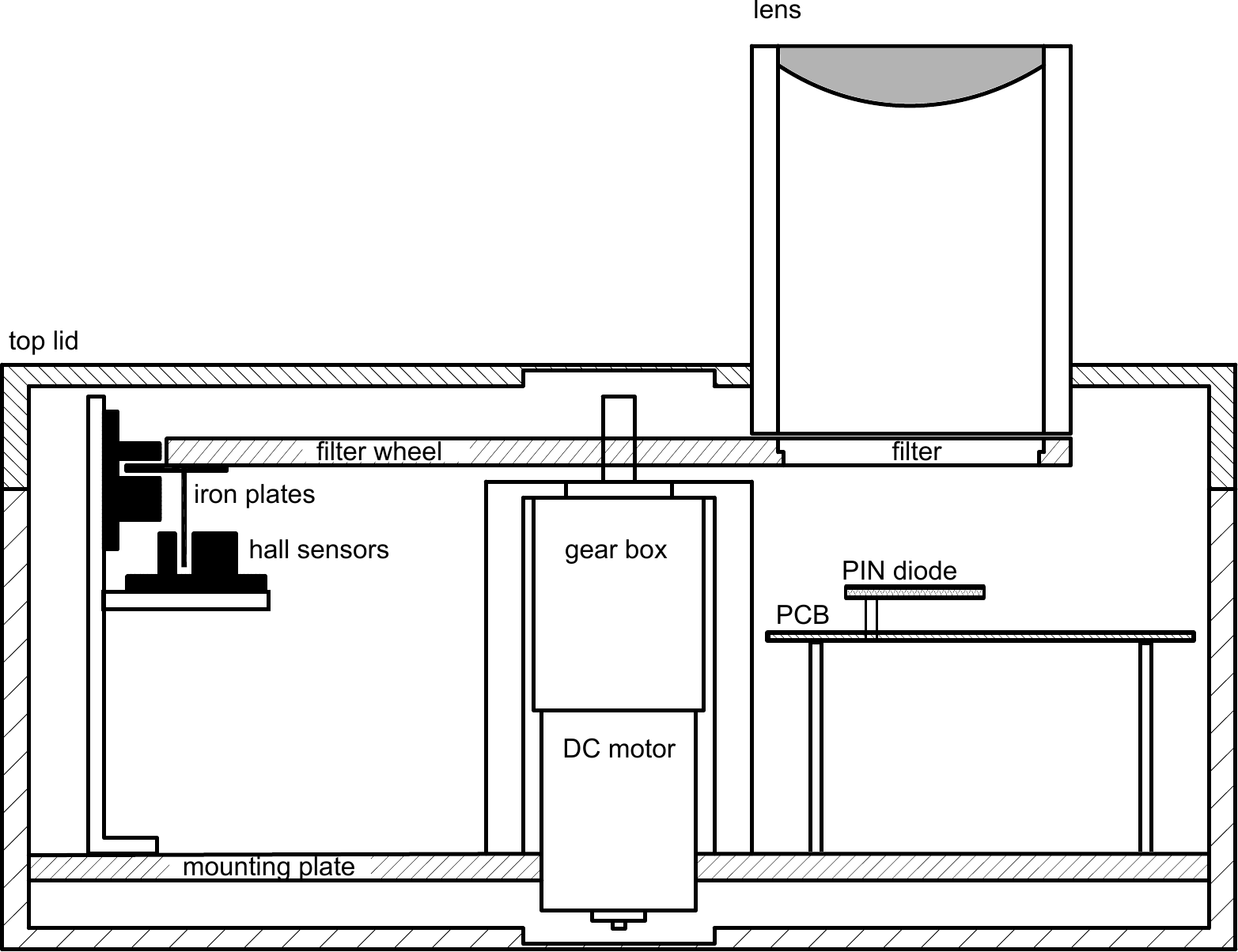}
	\end{center}
	\caption{Mechanical design of the LoNS sensor with its 5-position filter wheel.}
	\label{fig:filterwheel}
\end{figure}

To be able to measure the LoNS in different wavelength bands and for taking dark frames before each measurement, the instrument is equipped with a 5-position filter wheel (see Figure~\ref{fig:filterwheel}). 

%\begin{figure}
	%\begin{center}
		%\includegraphics[width=1.0\columnwidth]{img/fig_bfilter.pdf}
	%\end{center}
	%\caption{Transmission curves of the two filters, that were combined to receive a B-band like filter for the LoNS sensor. The measurements were performed in the QEMD box.}
	%\label{fig:bfilter}
%\end{figure}
\begin{figure}[h!]
\centering
\includegraphics[width=0.95\columnwidth]{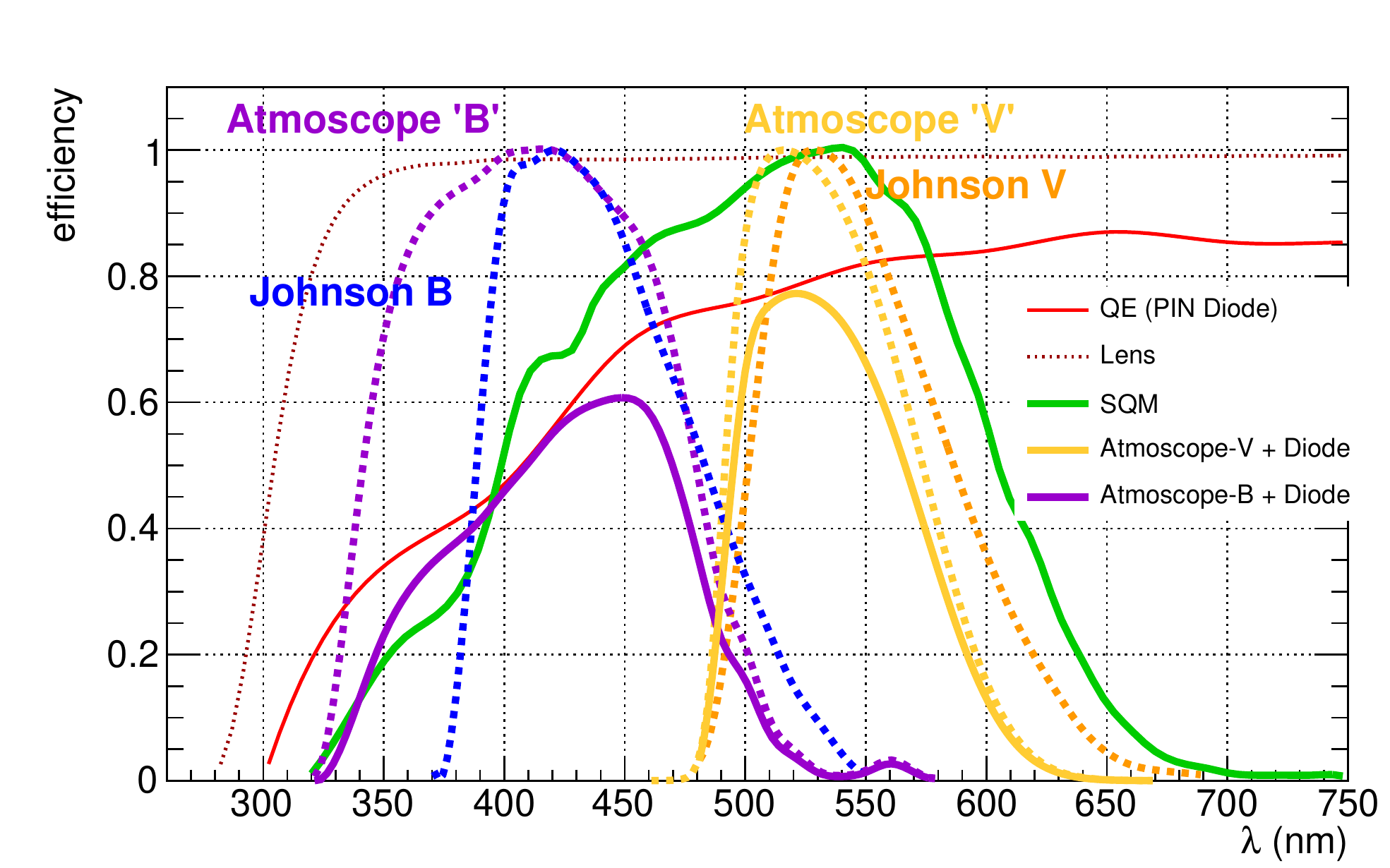}
\caption{Transmission curves of the Hamamatsu S3584-08 PIN diode (full red), the lens (dotted brown) and the ATMOSCOPE filters (violet and yellow). % and the atmospheric transmission for a typical clear astronomic night. 
The blue and orange lines show the standard Johnson/Bessell blue and visible filters for comparison. The dotted lines 
show the normalized transmission of the filters only, the full lines after convolution with the quantum efficiency of the PIN diode. % and the atmospheric transmission. 
For comparison, the normalized spectral acceptance of a Sky Quality Meter (SQM)~\protect\citep{cinzano} is shown in green.\label{fig:filters} }
\end{figure}

The 'closed' position enables dark frames to be taken. Another position is equipped with an astronomical V-band filter from Lot-Oriel~\citep{lotoriel}. The next position houses a set of two colour glass filters from Schott~\citep{schott} that in combination resemble an astronomical B-band filter but have an extension towards the UV. %See Figure~\ref{fig:filters} for the transmission curves. 
The remaining two positions are empty. 
%Figure \ref{fig:lons_box} shows the interior of the LoNS sensor with filter wheel removed.

Figure~\ref{fig:filters} shows the transmission curves for the filters used by the ATMOSCOPE. These 
match more-or-less the standard Johnson/Bessell visible and blue filters~\citep{johnson,bessell,bessell1}, but with some 
shift of the mean wavelength. The ``ATMOSCOPE-B''~filter moreover shows broader spectral acceptance in the UV than the standard blue filter (see Figure~\ref{fig:filters}) and hence covers 
the major part of the spectrum of Cherenkov light from air showers observed on ground. 
The effect of these shifts needs to be corrected for if star catalogues from standard filter photometry are used for calibration, or in order to compare results with other instruments. 
Peak transmission ranges from 0.85 to 0.88 for the visible band and from 0.61 to 0.71 in the blue/UV band. 
%For this reason, a \textit{reference box} has been chosen with respect to which all other filters have been calibrated. 
%That reference box has remained in the lab at CPPM. 

\begin{figure}[htp]
\centering
\includegraphics[width=0.99\columnwidth]{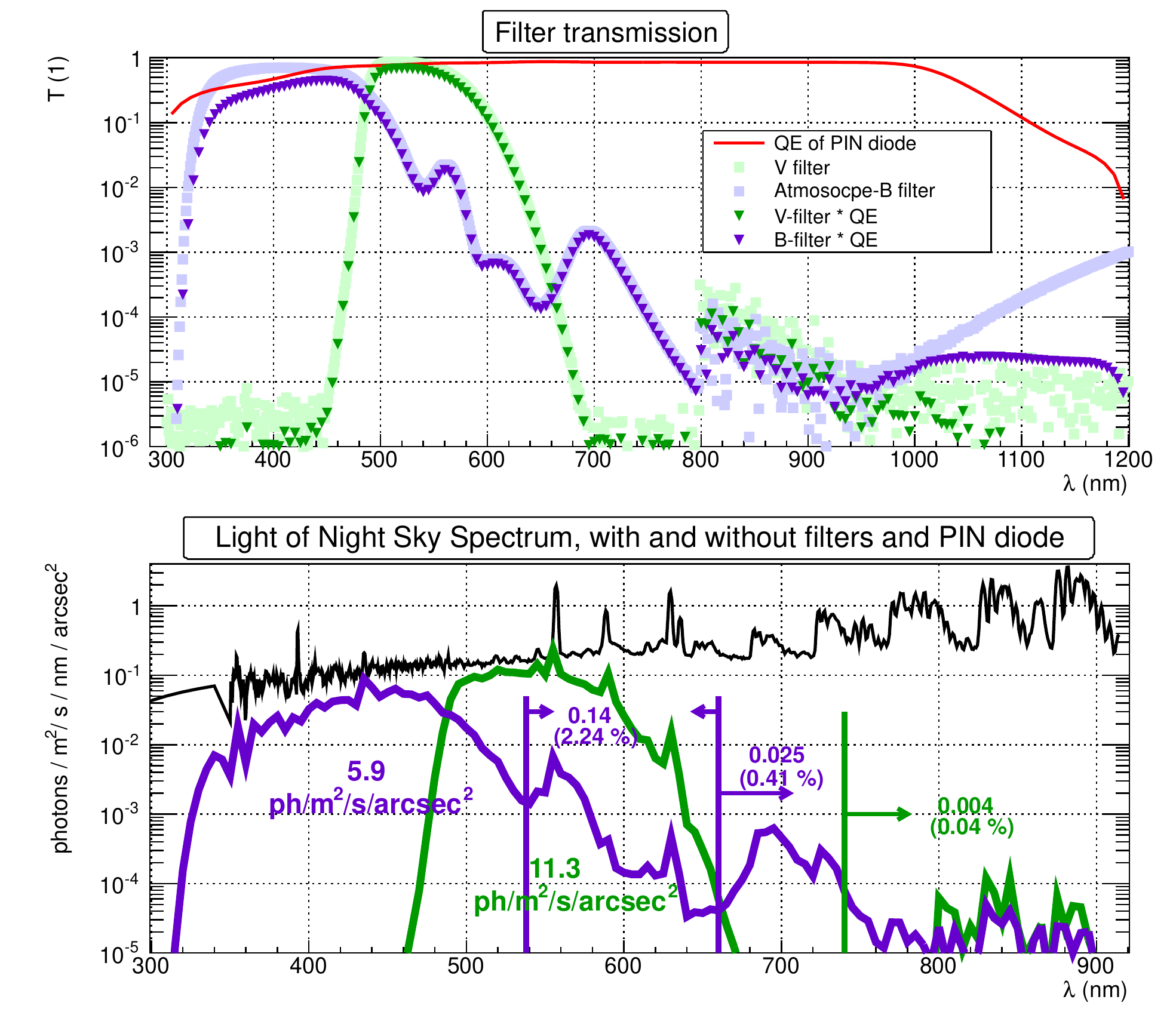}
\caption{V-filter and ``ATMOSCOPE-B'' filter leakages measured for an ATMOSCOPE produced at MPI. % (two plots above) and another produced at CPPM (two plots below). 
Shown are the transmissions, with and without 
folding with the quantum efficiency of the PIN diode, and the result folded with a typical light-of-night-sky spectrum \protect\citep[from][]{king}. The integral light-of-night-sky leakage 
into the V-filter window amounts to 0.04\%, the B-filter leakage into the V-filter window amounts to 2.2\% and into the infrared leakage to about 0.4\%. 
Measurements courtesy of Jose Luis Rasilla, IAC.\label{fig:leakage}}
\end{figure}

Special care has been taken to control 
filter leakage at other wavelengths, especially in the near infrared, where the night sky is much brighter (see Figure~\ref{fig:leakage}). 
The visible filter shows transmission values 
always below the $10^{-4}$ level outside the nominal window, and the cumulative contribution of leaking light from longer wavelengths can be conservatively estimated to be below 
0.1\,\% of the light registered inside the nominal window. The blue/UV filter shows stronger leakage, estimated to account for $\sim 3$\% 
for typical night-sky spectra~\citep{king}.
A small leak around 560~nm shows transmission up to $2\cdot 10^{-2}$, and coincides with the OI~557.7~nm emission line from airglow. 
Another broader leak allows $10^{-3}$ of the light to pass through at wavelengths around 690~nm. We conservatively estimate that 
the first leak may contribute up to 3\,\% of the nominal filter wavelengths, but this is well controlled by the measurements 
performed with the visible filter, which has overlapping acceptance. The red and near infrared leakage is smaller and estimated to contribute always less than 
1\,\%. 
%These numbers will be included in the systematic uncertainties of the measured residual light backgrounds, however they are not the limiting uncertainty.

\begin{figure}[h!]
\centering
\includegraphics[width=0.99\columnwidth]{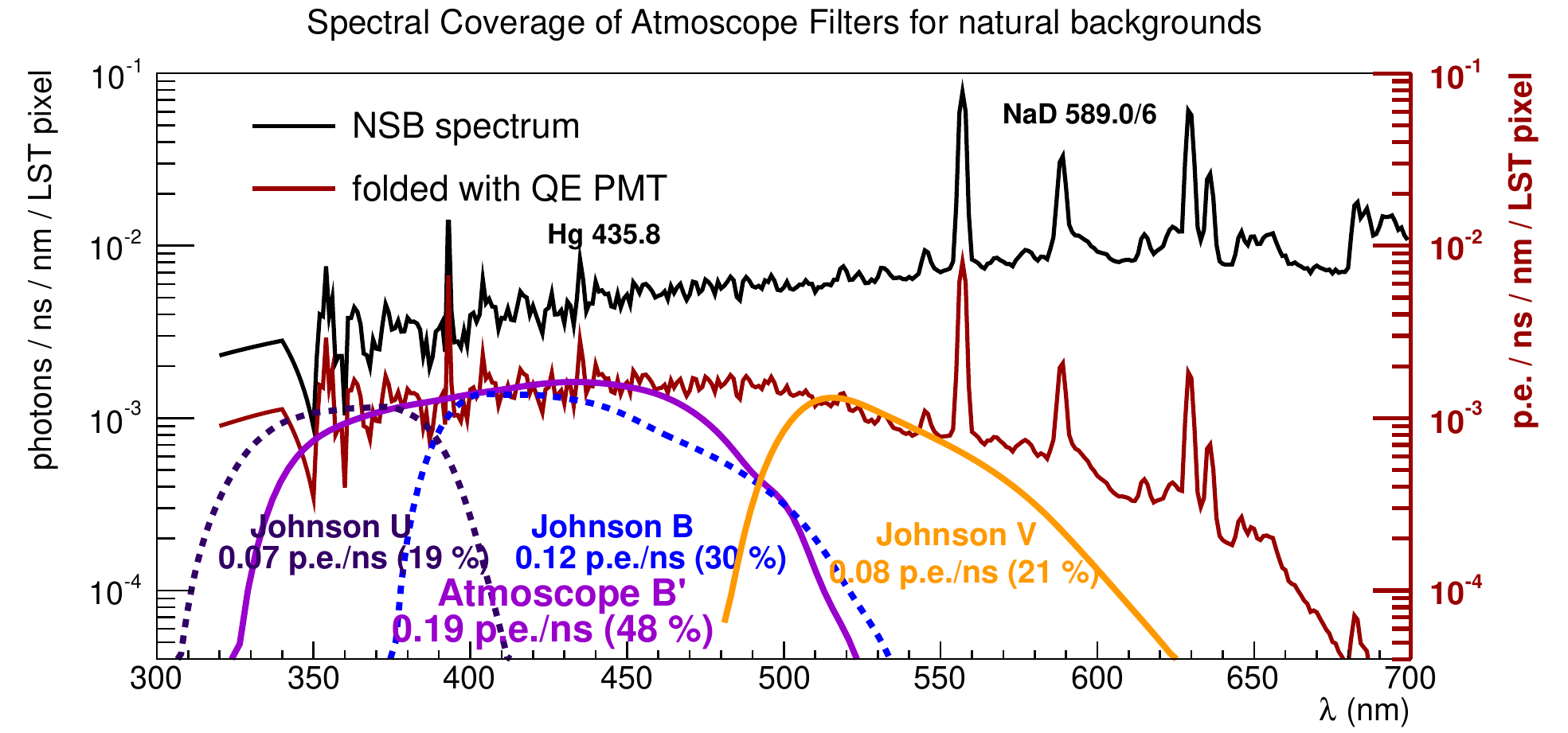}
\caption{Example of a night-sky spectrum from an astronomical site, here the Observatorio del Roque de los Muchachos at La Palma~\protect\citep{king}, 
in units of photons per nano-seconds, per nano-meter per pixel of the camera of the large-size telescopes of the CTA (black line). To a good 
approximation, anthropogenic backgrounds can be considered to be limited only to the two lines Hg$_{435.8}$ and NaD$_{589.0/6}$ for the ranges of interest here.
The brown line shows the 
same spectrum, folded with the quantum efficiency of the Hamamatsu S11920 PMT, used for the large-size telescopes of the CTA. 
%The corresponding scale (in photo-electrons per nano-seconds, per nano-meter and single pixel) is displayed on the right side. 
Also shown are the responses of the two filters used in the ATMOSCOPE, together with those of the two standard Johnson/Bessell 
visible and blue filters for comparison. The numbers below the filter curves show the integral in units of photo-electrons per nano-second, and their relative 
contribution to the observed photo-electron background.\label{fig:lonslst}}
\end{figure}

To demonstrate that the ATMOSCOPE incorporates a good choice of filters for the site selection of the CTA, Figure~\ref{fig:lonslst} shows the natural light of night sky spectrum, due to air-glow 
and the minimum zodiacal light~\citep{king}, folded with the quantum efficiency of photomultipliers foreseen for the 23~m Large-Size Telescope (LST)~\citep{teshima2013}.  
That telescope type has been chosen because its sensitivity will be limited most severely by the LoNS, compared to the Medium-Size ($\sim$12~m) and Small-Size ($\sim$5~m) telescopes.
One can see that the ``ATMOSCOPE-B'' filter covers 
almost half of the background light responsible for photomultiplier noise, while the V-filter covers about 20\%. There is only a small overlap between both filters, and the contribution 
of light from 600~nm onwards is largely suppressed by the spectral acceptance of the photomultiplier. Note that several cameras developed for the smaller telescopes of the CTA 
are planned to be equipped with Geiger-mode avalanche photo diodes (often referred to as ``Silicon PMs''), which show significant spectral acceptance up to 800~nm. However efforts 
are ongoing to cut out the (unwanted) acceptance by using filters in this range. 
%\begin{figure}[hbp]
%\centering
%\includegraphics[width=0.99\columnwidth]{img/LonsWeb.png}
%\caption{Example of a night-sky spectrum from an astronomical site, here the Observatorio del Roque de los Muchachos at La Palma~\protect\cite{king}, 
%folded with the spectral acceptance of the PIN diode and the two filters used in the atmoscope, and the two standard Johnson/Bessell 
%visible and blue filters for comparison. The numbers below the filter curves show the integral in units of magnitude per square arc-second.\label{fig:lons}
%\vspace{-5cm}}
%\end{figure}

\begin{figure}[htp]
\centering
\includegraphics[width=0.49\columnwidth]{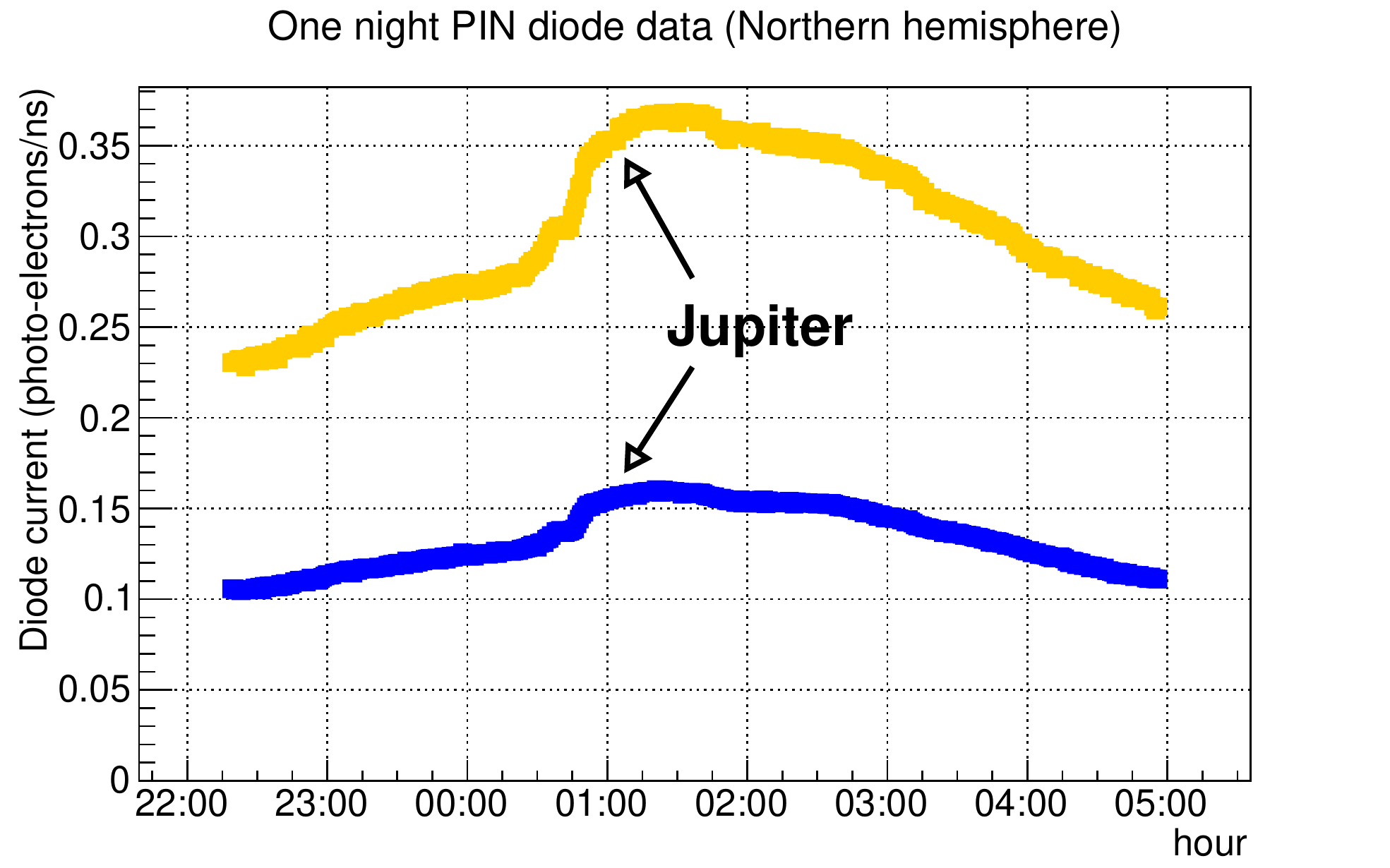}
\includegraphics[width=0.49\columnwidth]{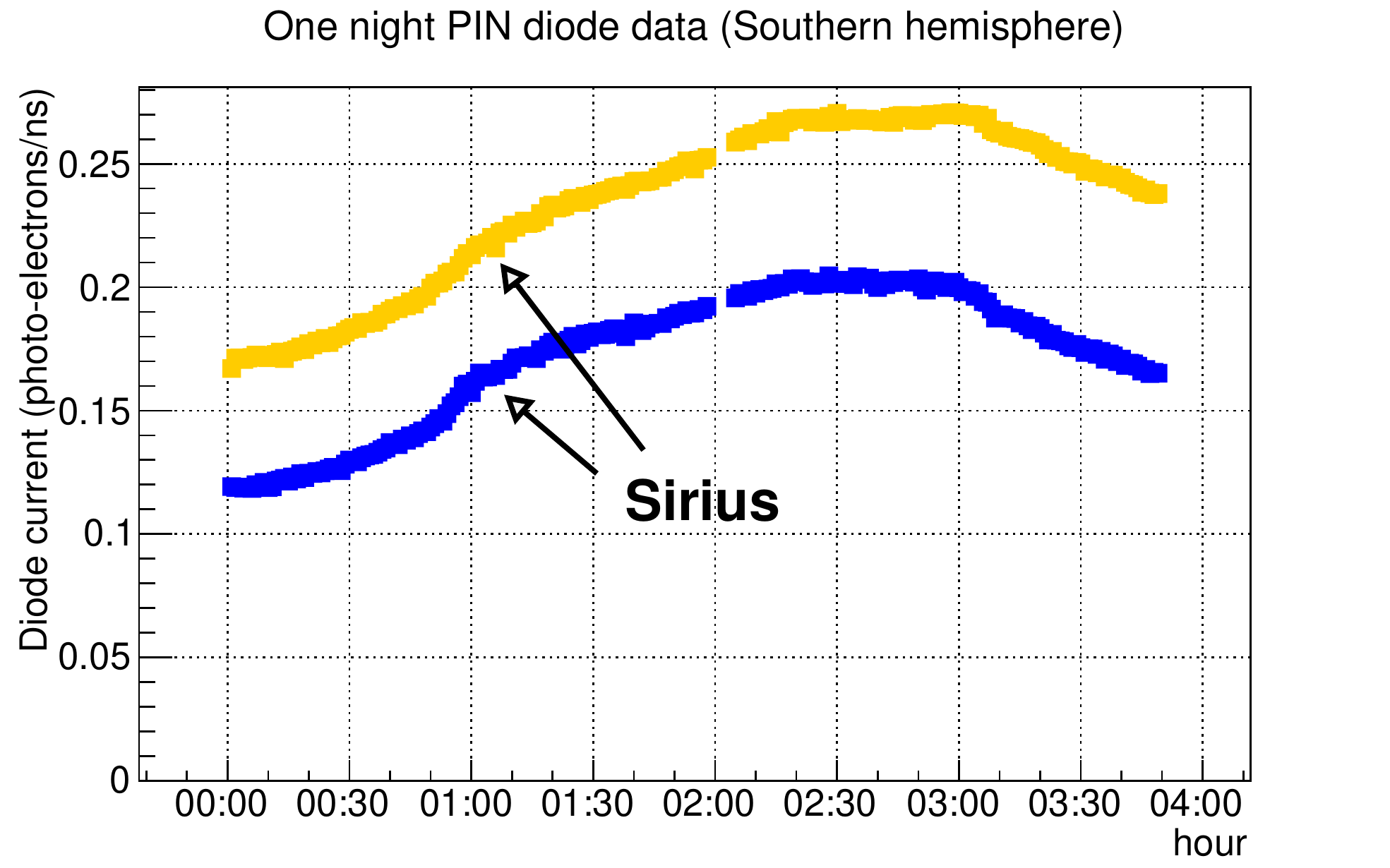}
\caption{Example of one night of night-sky measurements with both filters (yellow: V-filter, blue: ``ATMOSCOPE-B''-filter''). 
Left: Northern hemisphere with Jupiter entering the field of view, right: Southern hemisphere with Sirius entering the field of view. 
\label{fig:jupiter} }
\end{figure}

\begin{figure}[htp]
	\begin{center}
		\includegraphics[height=0.37\columnwidth]{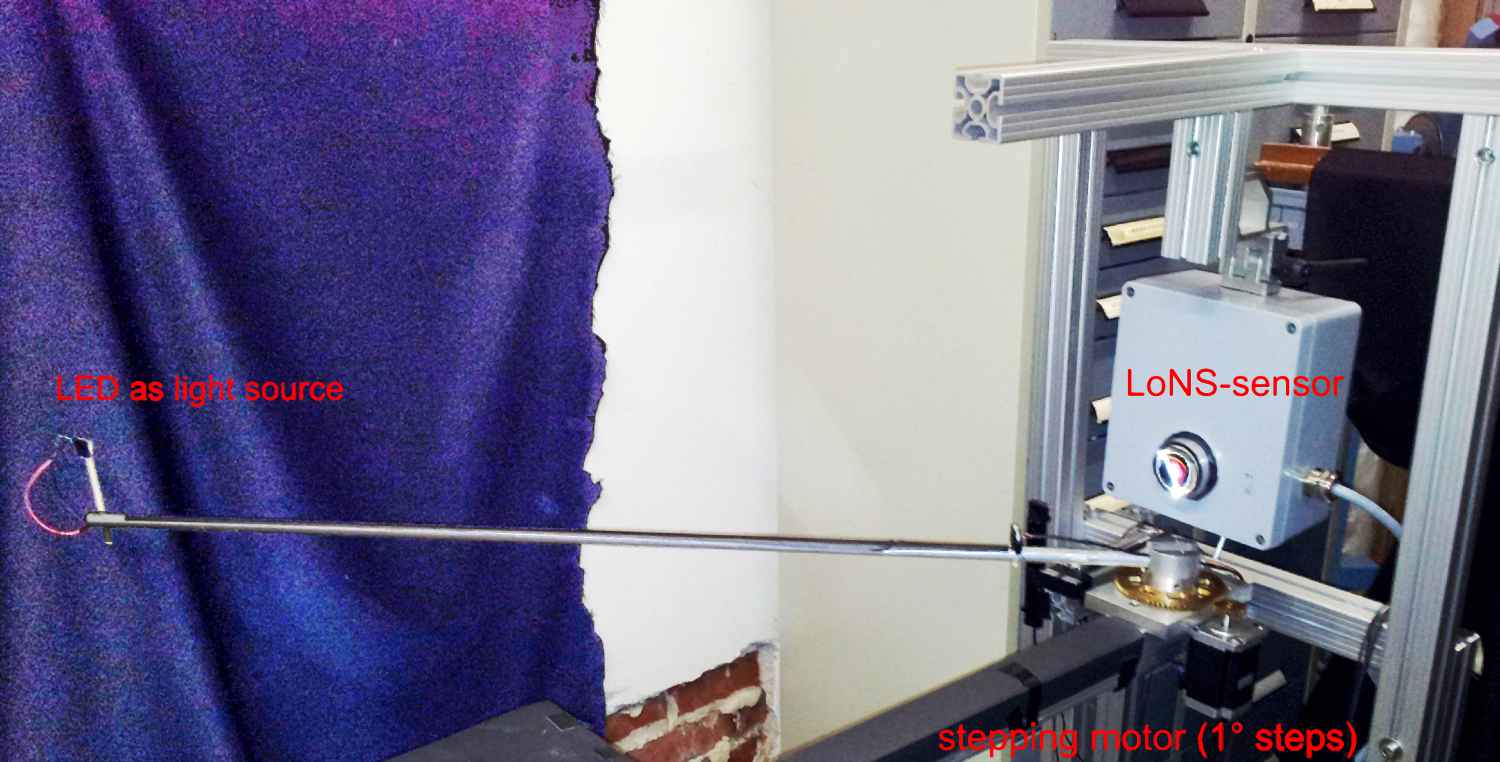}
		\includegraphics[height=0.37\columnwidth]{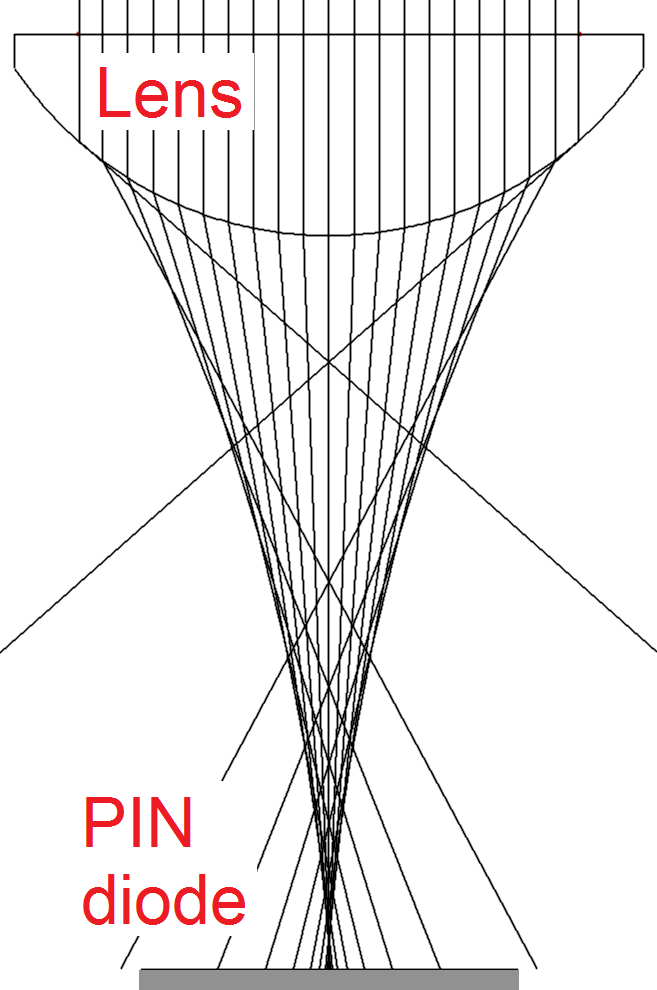}
	\end{center}
	\caption{Experimental setup for the acceptance measurement of the LoNS sensor (left). Ray-tracing simulation of the ATMOSCOPE lens for blue light using\protect\citep{winlens3d} (right).}
	\label{fig:acceptance}
\end{figure}

The lens of the LoNS device has a physical diameter of $\np[cm]{5}$ and a focal length of $\np[cm]{6}$. 
Its orientation has been chosen in order to produce a sharp cut-off at the edges of the acceptance function, such that bright point sources, like planets or bright stars, 
can be distinguished by a sudden change in the diode current when they enter the field of view. This feature was later used for fine-calibration of the effective orientation 
of the LoNS sensor (see Figure~\ref{fig:jupiter}).
With the PIN diode in its focal plane, the field of view is $\approx \np[sr]{0.38}$. Ray-tracing simulations (Figure~\ref{fig:acceptance}, right) show that not all the light collected on the surface of the lens is imaged onto the PIN diode. 
Assuming normal incidence, the maximum radius at which light does not undergo total reflection on the inside of the lens can be calculated using $R_\mr{max} = 1/n_c \cdot R_c$, where 
$R_\mr{c} = \np[mm]{30.9}$ is the radius of curvature of the lens, and the $n_\mr{c}$ the refractive index of glass, which can be approximated by $n_c \approx 1.52$ for the V-filter wavelengths, 
and $n_c \approx 1.53$ for the ``ATMOSCOPE-B''~filter. 
%	\begin{equation}
%		R_\mr{max} = \sin \left( \arcsin \left( \frac{1}{n_\mr{c}}\right) \right)\, R_\mr{c} = \np[mm]{20.394}
%	\end{equation}
%This results in a reduction of the available collection area for light by a factor $0.666$ compared to the geometric collection area. Because of the tails of the point spread function of the lens, the angular acceptance is not box shaped, but has tails as well. 
This results in an entrance pupil of 12.9--13.0~cm$^2$. 

\begin{figure}
\centering
\includegraphics[width=0.98\columnwidth]{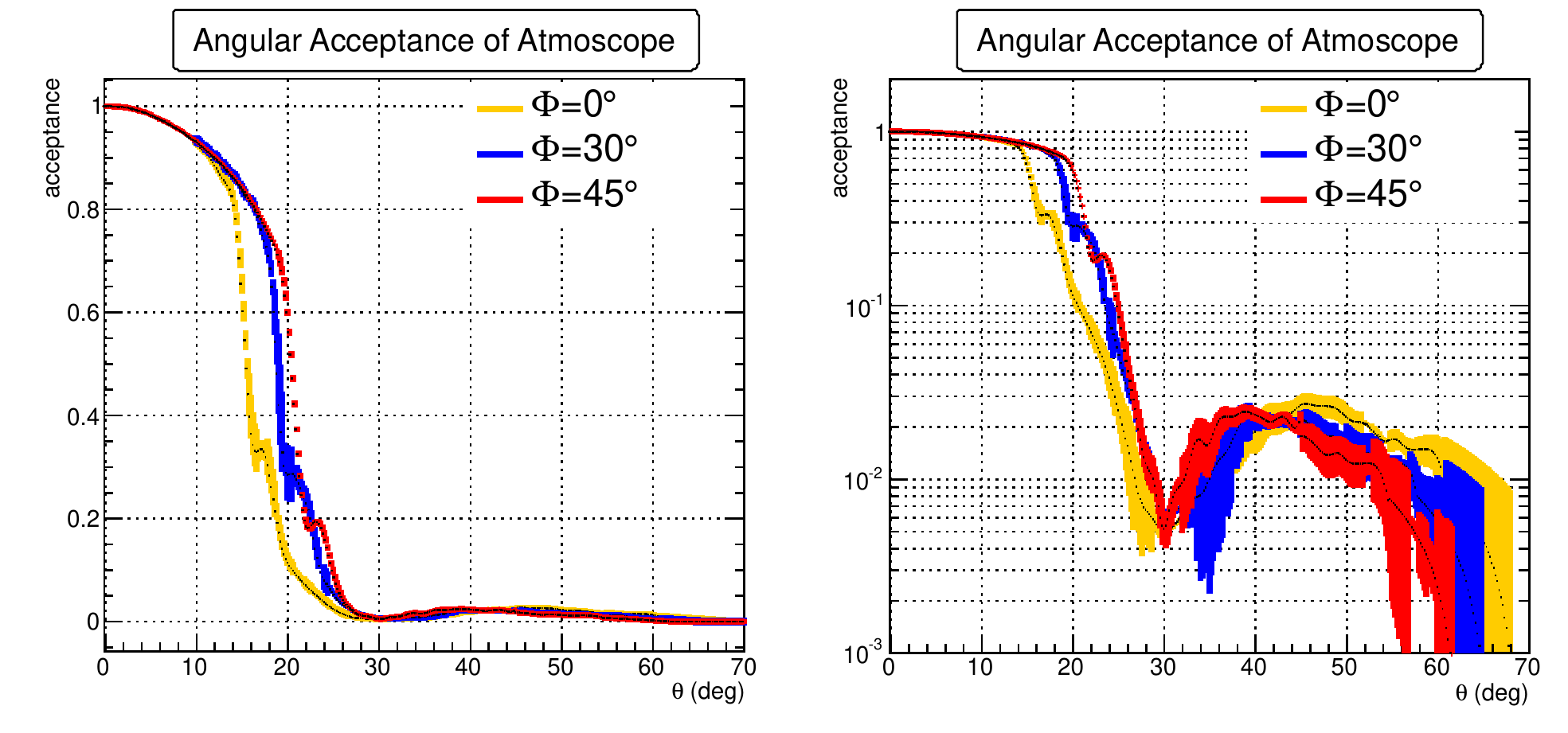}
\caption{Interpolated data from measurements at MPI of the angular acceptance of the ATMOSCOPE, as a function of the radial incidence angle $\theta$. The three curves show 
different azimuthal angles, where $\Phi=0$ corresponds to incidence on the side of the PIN diode, $\Phi=45$ along its diagonal. 
Left: linear scale, right: logarithmic scale. \label{fig:acceptancemeas}}
\end{figure}

Due to the square shape of the PIN diode, the acceptance of the light sensor is not symmetric in azimuth ($\Phi$). 
The angular acceptance of one ATMOSCOPE has been measured in the laboratory of the MPI Munich\footnote{Alternatively a circular diaphragm could have been used to make the acceptance rotationally symmetric. This way one would however have sacrificed some of the angular acceptance.}. 
This was done by scanning the radial incidence angle ($\theta$) in $\np[\deg]{1}$ steps for 3 different $\Phi$ positions, 
$\np[\deg]{0}$, $\np[\deg]{30}$ and $\np[\deg]{45}$ (see Figure~\ref{fig:acceptance}). 
Due to mechanical constraints in the setup, the 
measurements were not extended to zenith angles larger than $\np[\deg]{60}$.
%, since no acceptance was expected there, due to the aperture-limiting lens. 
Values between $\np[\deg]{60}$ and $\np[\deg]{70}$ were obtained through extrapolation.
%the results from 55$^\circ$ to 60$^\circ$ have been extrapolated by 10 more degrees to about 70$^\circ$ in total.
The resulting acceptance is symmetric with respect to the sign of the zenith angle %, the remaining azimuth angles were interpolated between the measured ones 
(see Figure~\ref{fig:acceptancemeas}). One can see that the acceptance drops down to less than 1\% at 30$^\circ$ incidence angle, as expected, but it rises again to reach 
a secondary maximum of 3\% around 45$^\circ$ incidence angle. Although 3\% seems a small number, the enhanced solid angle between 35$^\circ$ and 60$^\circ$ 
makes this contribution significantly larger than 10\%.
The secondary acceptance can be explained with single scattering of light on the surface of the ring which holds the lens. 
Although the ring is painted with black anti-reflective paint, a 3\% reflectivity is quite possible. 
Any tertiary acceptance due to 
multiple scattering should be suppressed by at least a factor $0.03^2 \sim 10^{-3}$, and is hence expected to contribute only insignificantly to the overall acceptance.

\begin{figure}
\centering
\includegraphics[width=0.98\columnwidth]{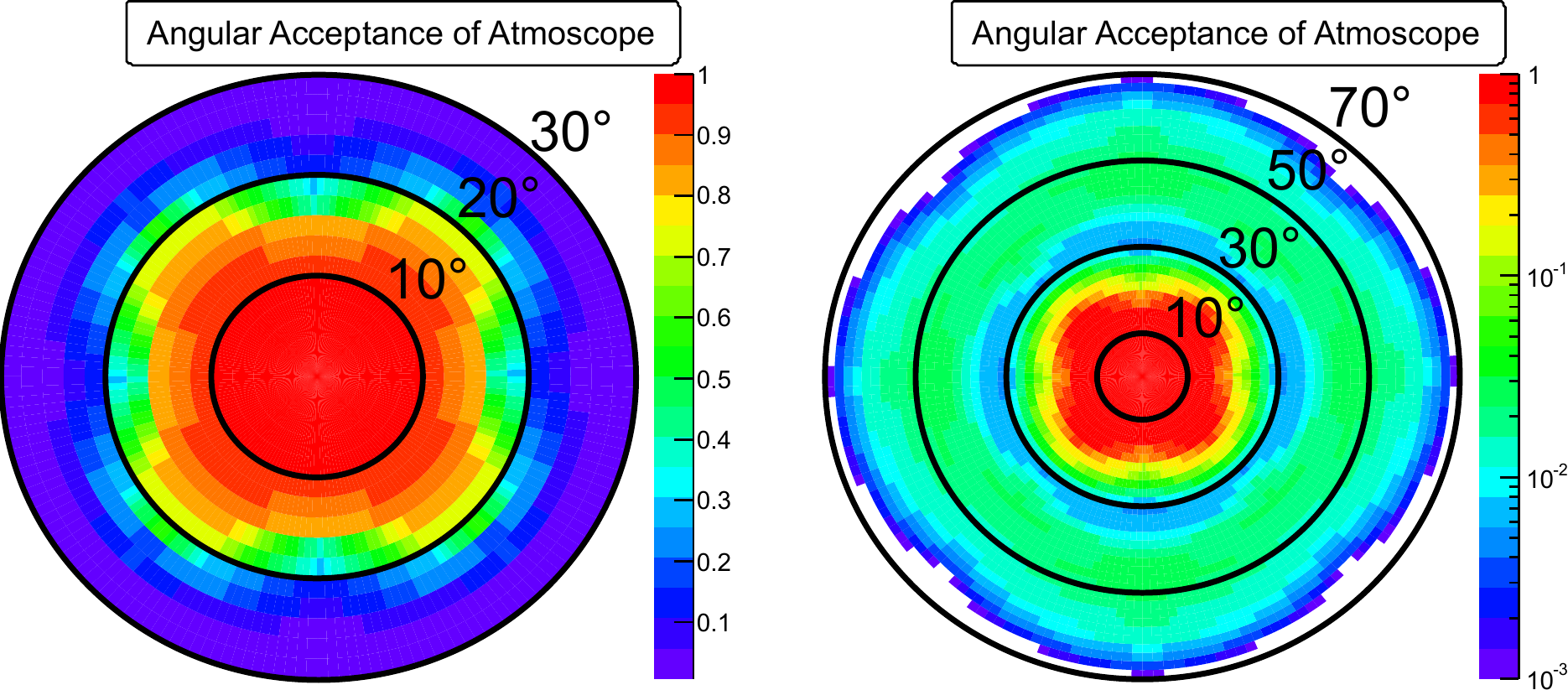}
\caption{Angular acceptance of the ATMOSCOPE, shown in polar coordinates up to an incidence angle of 30 degrees (left, linear scale) and up to 70~degrees (right, logarithmic scale). 
\label{fig:acceptancepolar}}
\end{figure}

A perfect 8-fold symmetry of the system was then assumed, and the statistical measurement uncertainties doubled at every 10$^\circ$ distance from the closest azimuth measurement. 
Even so, the statistical uncertainties of the overall measurement of the cumulative angular acceptance are less than 1\%.
The result of this analysis is shown in Figure~\ref{fig:acceptancepolar}.
These measurements were made with a blue LED, and hence no spectral dependency of the acceptance could be measured. However, given the small variation of the refractive index of the
lens glass with wavelength, this effect is likely limited to less than 3\%.

\begin{figure}[h!]
\centering
\hspace{-1cm}
\includegraphics[width=0.83\columnwidth]{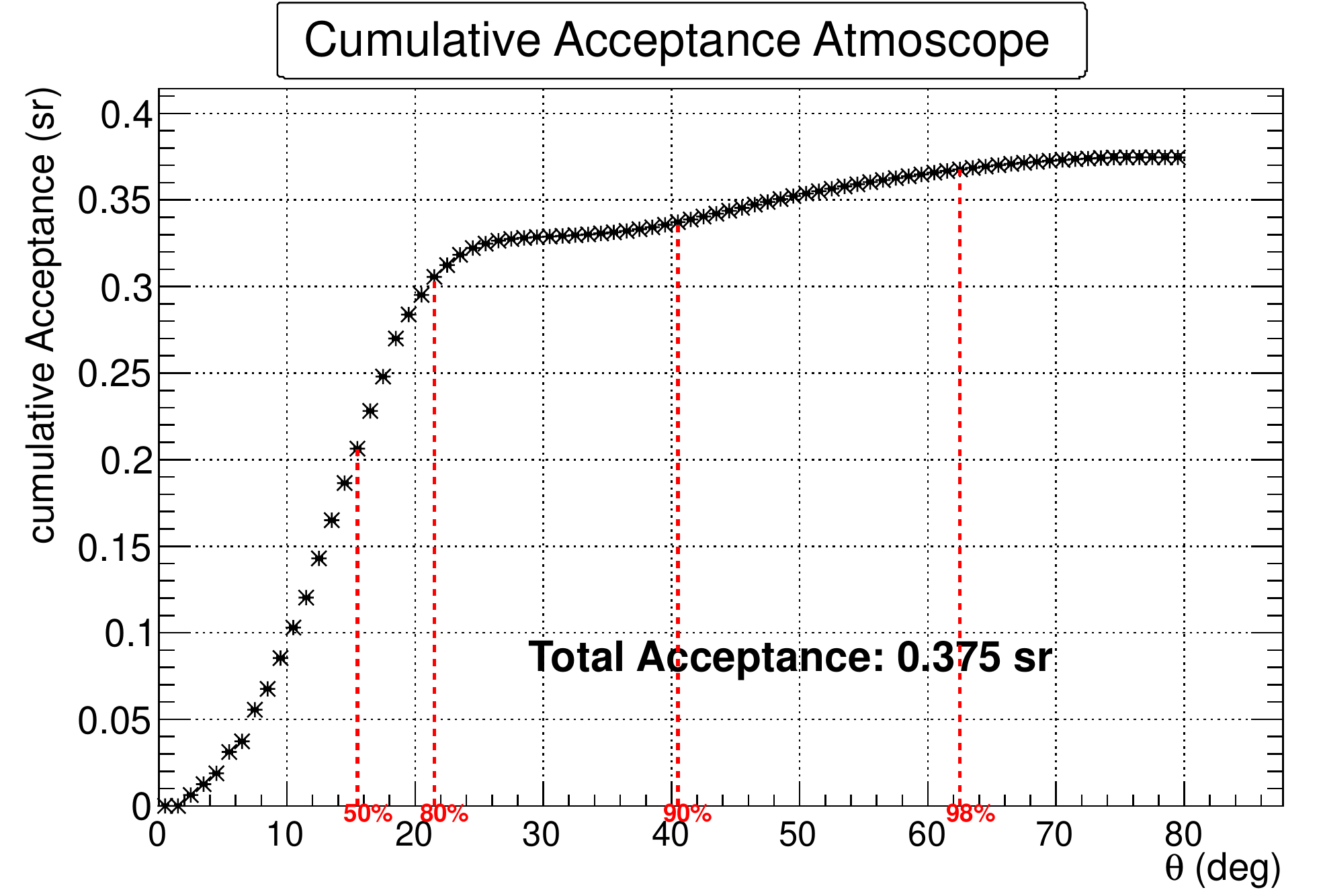}
\caption{Cumulative acceptance from the lab. measurements (see Figure~\protect\ref{fig:acceptance}). 
The red lines denote those angles where 50\%, 80\%, 90\% and 98\% of total acceptance is reached. 
%The error bars reflect statistical measurement uncertainties only. 
For zenith angles larger than 70$^\circ$, the acceptance has not been measured, but can 
be assumed to contribute by another 0.007~sr. \label{fig:acceptancecum}}
\end{figure}

The resulting cumulative angular acceptance is 0.375$\pm 0.005$~sr (see Figure~\ref{fig:acceptancecum}). 
The statistical and systematic uncertainty stems almost entirely from the extrapolation from 60$^\circ$ to 70$^\circ$. 
%, assuming perfect 8-fold symmetry of the system and doubling of the measurement uncertainty at every 10 degrees distance 
%to the closest azimuth measurement. 
Degradations of the acceptance due to dirt and aging are not included here, and have been calibrated on-site 
about once per two months.

The three analogue PIN diode current measurement channels are read by a MAXIM~128 ADC, with a range of 0 -- $\np[V]{4.095}$ and a resolution of $\np[mV]{1}$. 
The communication to the control computer is via I$^2$C, a simple serial protocol, using one data line and one clock line.
There is a second optional temperature channel that can be used to monitor the temperature inside the electric control cabinet. 
The humidity inside the LoNS sensor box is also monitored by a sensor connected to the MAXIM 128. 
	
The custom-made filter wheel is turned by a DC gear motor which is switched on and off with a PhotoMOS relay from the computer. 
The positioning feedback comes from two Hall sensors mounted on the filter wheel, that detect the iron plates passing through.
One sensor gives the computer the stop signal at each of the five positions,  
and the other, which is mounted at the position of the dark measurement (taken before each run) provides absolute positioning.
The diode and the electronics are installed inside a water proof box, with the lens elevated slightly above the cover, positioned at the end of a tube. 

The average LoNS on a dark site between 300 and 600~nm is about $\np[ph \ m^{-2} sr^{-1} s^{-1}]{1.7E12}$~\citep{razmikhegra}. 
Assuming an average QE of about $\np[\%]{50}$ for the ``ATMOSCOPE-B''~filter, an average filter transmission of about 0.65, 
the collection area of the lens of $\np[m^{2}]{1.3E-3}$ and a field of view of $\np[sr]{0.38}$, roughly
$\np[ph \ s^{-1}]{2.7E8}$ would be expected for the ATMOSCOPE to be measured under dark sky conditions. 
The corresponding PIN diode produces about 0.27 photo-electrons per nano-second which yields a 
diode current of $\np[pA]{43}$.
In the three different readout channels with amplification of ($\np[VA^{-1}]{E8}$, $\np[VA^{-1}]{E9}$ and $\np[VA^{-1}]{E10}$), 
this gives signals of $\np[mV]{4.3}$, $\np[mV]{43}$ and $\np[mV]{430}$, respectively. 

The PIN diode itself produces a temperature-dependent dark current, 
even in the absence of light. Since it is operated in photo-voltaic mode, the PIN diode will reach a saturation reverse current which can be described by~\citep[see e.g.][]{sze}: 
% Formula 46, page 102
\begin{eqnarray}
I_s &=& A \cdot \frac{D_p\cdot c_n^2}{L_p \cdot N_D} \cdot T^{(3 + \gamma/2)} \cdot \exp(-E_g/k T) \nonumber\\
    &\approx& \frac{10^{11}~\mathrm{ns^{-1}~K^{-3}}}{(L_p/\mathrm{100\,\mu m}) \cdot (N_D/\mathrm{10^{21}\,m^{-3}})} \cdot{}
    T^{(3 + \gamma/2)} \cdot \exp(-E_g/k T) 
\end{eqnarray}
\noindent
$A = 7.84\cdot 10^{-4}$~m$^2$ is the PIN diode area. $D_p = 1.2\cdot 10^{-3}$~m$^2$/s is the diffusion coefficient of holes in silicon. $c_n
= 3.28\cdot 10^{21}$~m$^{-3}$~K$^{-3/2}$ is the effective density of states in the un-doped layer. $L_p$ is the diffusion length of holes, and $N_D$ 
is the donor density. The additional exponent $\gamma$ parameterizes the temperature dependency of the diffusion length. Finally, $E_g = 1.12$~eV is
the band gap energy of silicon, and $k$ the Boltzmann constant.
% where $A\approx 7.84\cdot 10^{-4}$~m$^2$ the PIN diode area, $D_p \approx 1.2\cdot 10^{-3}$~m$^2$/s the diffusion coefficient of holes in silicon, $c_n 
% \approx 3.28\cdot 10^{121}$~m$^{-3}$~K$^{-3/2}$ the effective density of states in the un-doped layer, $L_p$ the diffusion length of holes, and $N_D$ the donor density. 
% The additional exponent $\gamma$ parameterizes the temperature dependency of the diffusion length. Finally, $E_g \approx 1.12$~eV is 
% the band gap energy of silicon, and $k$ the Boltzmann constant. 

\begin{figure}
	\begin{center}
		\includegraphics[width=0.75\columnwidth]{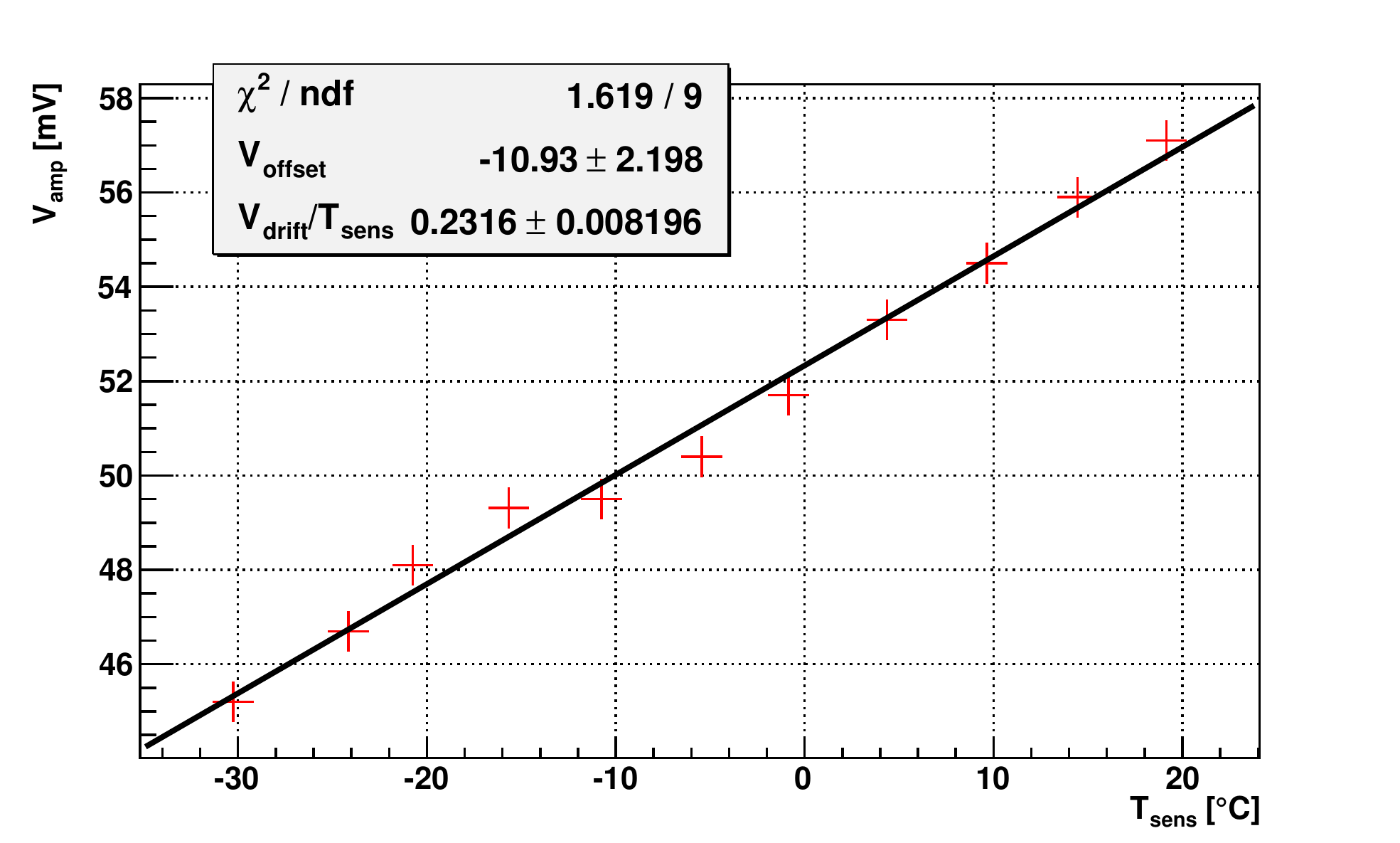}
	\end{center}
	\caption{%
%Temperature dependence of the dark current through the PIN diode. The upper graph shows the measured voltage signal after the amplifier, measured in the climate chamber of the institute, at temperatures between $\np[\deg C]{-30}$ and $\np[\deg C]{30}$. The theoretical prediction for the diodes conductivity plus two terms that model the temperature dependent offset of the amplifier, are fitted to the data. To check the amplifier offset and the linear temperature drift, also the empty amplifier was measured. 
Our measurement of the linear temperature drift of the empty AD549 amplifier (diode not plugged in).  }
%The temperature drift $V_{drift}/T_{sens}$ in the lower graph is in good agreement with the measurement with diode. The temperature enters the fits in Kelvin.}
%	\label{fig:tempdiode}
\label{fig:tempdrift}
\end{figure}

The saturation current $I_s$ is expressed here in units of electrons per nano-second. 
Additionally, the amplifier shows a linear temperature drift (see Figure~\ref{fig:tempdrift}).
%, guaranteed to be less than $3\cdot 10^{-4}$~ns$^{-1}$K$^{-1}$ by the manufacturer.   
A small voltage offset was chosen by design, in order to ensure always a positive dark current measurement.

\begin{figure}[h!]
\centering
\includegraphics[width=0.49\columnwidth]{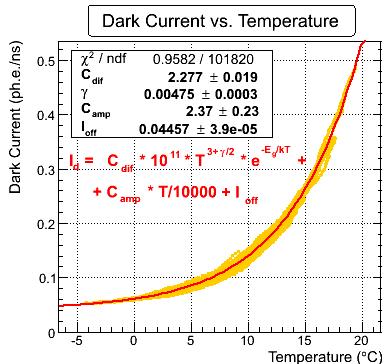}
\includegraphics[width=0.49\columnwidth]{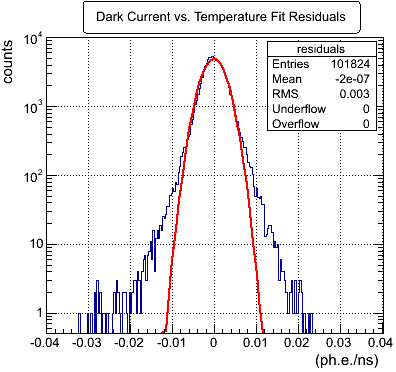}
\caption{Left: The measured mean dark current (expressed in photo-electrons per nano-second) vs. the temperature 
measured at the PC-board and fitted to diffusion dark current contribution, an amplifier bias 
offset drift, and a constant offset (red line). The measurement points correspond to a one-year measurement on a site with relatively small temperature gradients.
\label{fig:darkvstemp}}
\end{figure}

Figure~\ref{fig:darkvstemp} (left) shows the measured dark currents as a function of temperature of one ATMOSCOPE, fitted to 
appropriate temperature dependencies for the PIN diode, amplifier bias current and offset. The measurement points were taken continuously from dark measurements throughout about one year. 
The fit describes the data well, as can be seen in the fit residuals plot on the right side. Less than 4\% of the fit residuals lie outside the fitted Gaussian;
this slight deviation may be explained by circumstances in which the PIN diode has a slightly different temperature than the sensor on the PC board.
Fitting all 9 deployed ATMOSCOPEs with the combination of these dependencies, we obtain $(2.3\pm 0.2)$ for the term $1/\big((L_p/\mathrm{100\,\mu m}) \cdot (N_D/\mathrm{10^{15}\,cm^{-3}})\big)$, 
which is compatible with the assumption that $L_p$ is about the same size as the depletion region, i.e. $\sim$30~$\mu$m, and an $N_D$ of the order of $\mathrm{10^{15}\,cm^{-3}}$. 
The fits yield results for the additional exponent $\gamma$ always smaller than 0.014, but larger than 0.0002. Finally, the temperature drift of the amplifier contributes with 
$\ud I_s(\mathrm{amp})/\ud T  = (7\pm4) \cdot 10^{-5}$~ns$^{-1}$K$^{-1}$.

%The measurements have been fitted to a \textit{diffuse dark current contribution}, 
%multiplied with the parameter $C_\mathrm{dif}$, a bias current drift from the amplifier, 
%multiplied with the parameter $C_\mathrm{amp}$, and a constant offset $I_\mathrm{off}$, adjusted once, prior to installation. 
%No \textit{depletion dark current} contribution could be detected. 
%The following PIN diode parameters have been estimated: electron diffusivity: $D_n=25$~cm$^2$/s, PIN diode area $A=7.84$~cm$^2$, $c_n=3.284\cdot 10^{15}$~cm$^{-3}$~K$^{-3/2}$, 
%characteristic length $x_c=30\mu$m, acceptor concentration $N_A=3\cdot 10^{15}$~cm$^{-3}$. 

The temperature dependency of the dark current is hence well understood, and always meets 
the specifications of the manufacturer. On the other hand, fluctuations around the fitted dependencies are still too large to predict the dark current with sufficient
precision using temperature measurements alone. For this reason, care was taken to measure the dark current with sufficient precision a few seconds before each 
filter measurement. This is achieved by putting the filter wheel into the completely closed position for recording the dark current before each LoNS measurement 
is made with the filters.
At each position the ADC is read 1000 times and the average value as well as the RMS are stored. The measured average dark current is immediately subtracted from the filter 
measurements, and stored in the data for cross-checks. 
%, but was not written to disk for the early ATMOSCOPEs (\textit{HESS-I, Aar-I,}, first part of \textit{Leoncito} and \textit{Teide-I}).

%The following plots demonstrate the stability of the dark current measurements. First of all, the fitted {\bf diffusion dark current} 
%component and the {\bf bias offset drift} of the amplifier are subtracted from the dark current measurements, in order to get rid of the (strong) temperature 
%dependencies. The remaining offsets are plotted against time in Figures~\ref{fig:darkvstimeS} and~\ref{fig:darkvstimeN}). At very few times, 
%small jumps in the dark current are seen. Some of them can be explained by hardware changes (e.g. La Palma), others are not yet understood. 

We define the signal-to-noise ratio (SNR) as the background subtracted filter current, divided by the statistical uncertainty of the measurement:

\begin{equation}
\textrm{SNR}_\mathrm{B,V} = \frac{I_\mathrm{B,V} - I_\mathrm{dark}}{\sqrt{\textrm{RMS}(I_\mathrm{dark})^2+\textrm{RMS}(I_\mathrm{B,V})^2}} \qquad. \label{eq:snr}
\end{equation}

The obtained SNR is always above 50 for the V-filter measurements, and above 30 for the ``ATMOSCOPE-B''~filter, even in the absence of anthropogenic light. 
This is enough to reduce statistical fluctuations of all measurements to well below the systematic uncertainties. 

Even in one case of a faulty high-gain channel, the SNR obtained from the medium gain always yielded values greater than 8.

%The temperature of the PIN diode is measured by a temperature sensor attached to its back side that is also read by the ADC, to be able to estimate the dark current (Figure~\ref{fig:tempdiode}), 
%that comes from the amplifiers input offset voltage. 
%This dark current depends on the diodes conductivity that is given by the number of electron hole pairs in the conductivity band at a given temperature and the amplifiers input offset voltage.
%		\begin{equation}
%			I_{dark} = A \ T^3 \exp{\left( \frac{E_g}{k_B \ T} \right)}
%			\label{eq:dark_current}
%		\end{equation}
%		Here, $A$ is a proportionality constant, $T$ is the temperature of the semiconductor, $E_g$ its band gap and $k_B$ the Boltzmann constant. 
%	\color{black}
	
%%The sensitivity of the sensors, and especially the lenses, is monitored from time to time with a calibration device emitting continuous light 
%at various, controlled intensities (the so-called ``Malus''-device). 
%Typically, a calibration is performed directly before and after cleaning the lens, and the difference attributed 
%to dust deposits on the lens. Then, a linear interpolation is made in time, between the reduced sensitivity from before cleaning, 

 \section{Calibration of the Light of Night Sky sensor}

       \begin{figure}[h!]
               \begin{center}
                       \includegraphics[height=0.45\columnwidth]{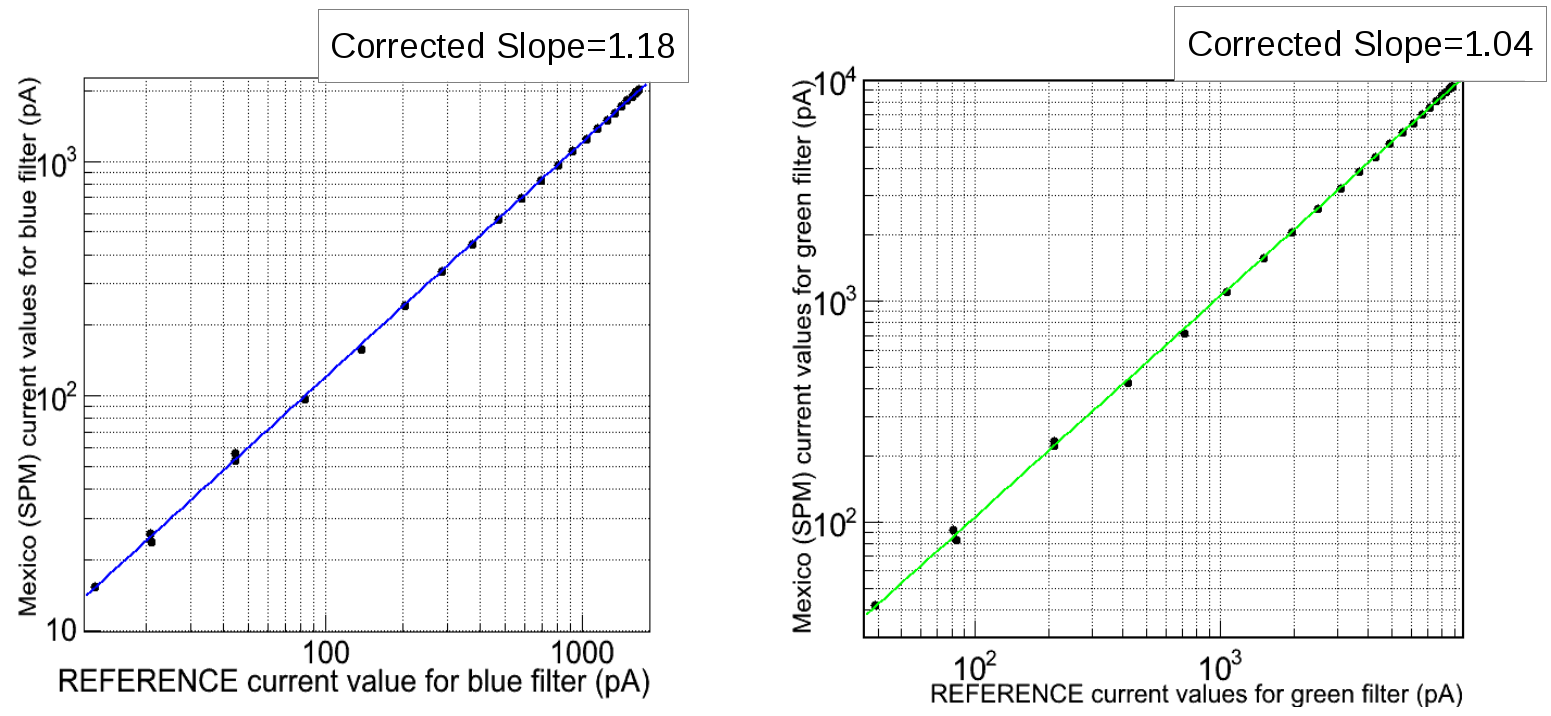}
               \end{center}
               \caption{Calibration of the LoNS sensors in the laboratory, before deployment, using continuous light and polarizers. The left plot shows the result of the ``ATMOSCOPE-B''-filter channel
                of one of the LoNS sensors
                (later sent to a candidate site) versus the reference LoNS sensor, which was kept in the lab. The right plot shows the same cross-calibration, for the V-filter.
		The slopes are corrected to take into account the change of the LED temperature and LED current which may evolve slightly 
		between the measurement of the two LoNS sensors.}
               \label{fig:maluscalibrationcurves}
       \end{figure}

\begin{figure}
               \begin{center}
                       \includegraphics[width=0.67\columnwidth]{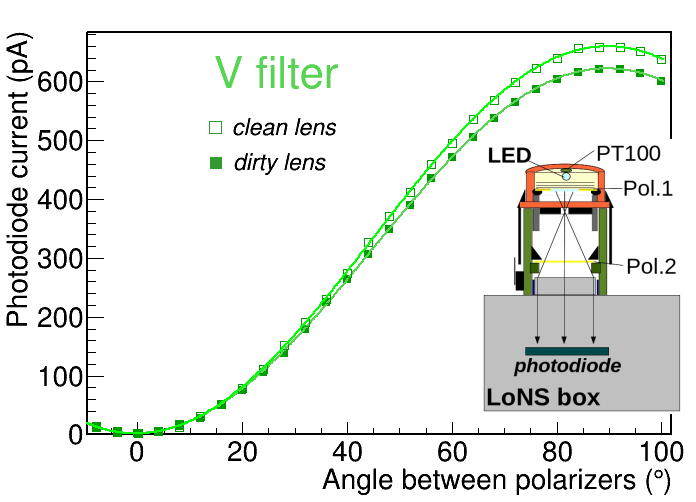}
               \end{center}
               \caption{Example of measurement of the ``dirtiness'': the response of the LoNS sensor to a tunable light source is measured 
	        before and after cleaning the lens. The ratio between the two response curves yields the ``dirtiness'', which is 5\% in this example.
		The layout of the calibration device is also shown. $Angle=0$ corresponds to crossed polarization directions.}
               \label{fig:maluscalibrationexample}
       \end{figure}

 Before the deployment of the devices a cross-calibration was performed in the laboratory.
 A reference light source made of a continuous, wide-spectrum LED and two polarizers was used for this purpose. Tuning the angle between the two 
polarizers allows for variation of the level of light according to Malus law for measuring the response of a LoNS sensor, and for checking the linearity across at least two gain ranges.
%polarizers allows 
 %the level of light to be varied according to Malus' law and allows for the measurement of the response of a LoNS sensor over a wide range of illumination and
 %for linearity checks across at least two gains. 
% The light emitted is continuous.
 Examples of calibration curves are shown Figures~\ref{fig:maluscalibrationcurves} and~\ref{fig:maluscalibrationexample}, the
 latter presenting the layout of the device.  The knowledge of the absolute level of light generated by the device
 is not necessary, as this apparatus aims at providing a 
reproducible set of light intensities over a wide range to compare the responses of different LoNS sensors. The 
 ratio of the measured responses gives the cross-calibration factor, which covers the whole system: lens, filters, photo-diode, and
 readout chain.
The accuracy of a cross-calibration between two sensors is of the order of $\pm$2\% at same LED current and temperature.

 One of the LoNS sensors has been kept in the laboratory and has served as reference 
 for calibrations: each LoNS sensor has been calibrated relative to this reference.
 Figure~\ref{fig:maluscalibrationcurves} shows the response of a sensor which was later sent to a candidate site versus the response of the 
 reference sensor: a difference of 18\% is measured for the response with the ``ATMOSCOPE-B'' filter, while only a 4\% difference is measured for the V filter. 
 This is consistent with the observation that the spread between batches of Schott B-filters was 
 larger than for the Lot Oriel V-filters. 
 Of the 9 LoNS sensors, the cross-calibration factors w.r.t. the reference sensor range from 0.96 to 1.07 when the V filters are in place, and from 1 to 1.2
 with the ``ATMOSCOPE-B'' filters, respectively. 
       
\begin{figure}
\centering
\includegraphics[width=0.85\columnwidth]{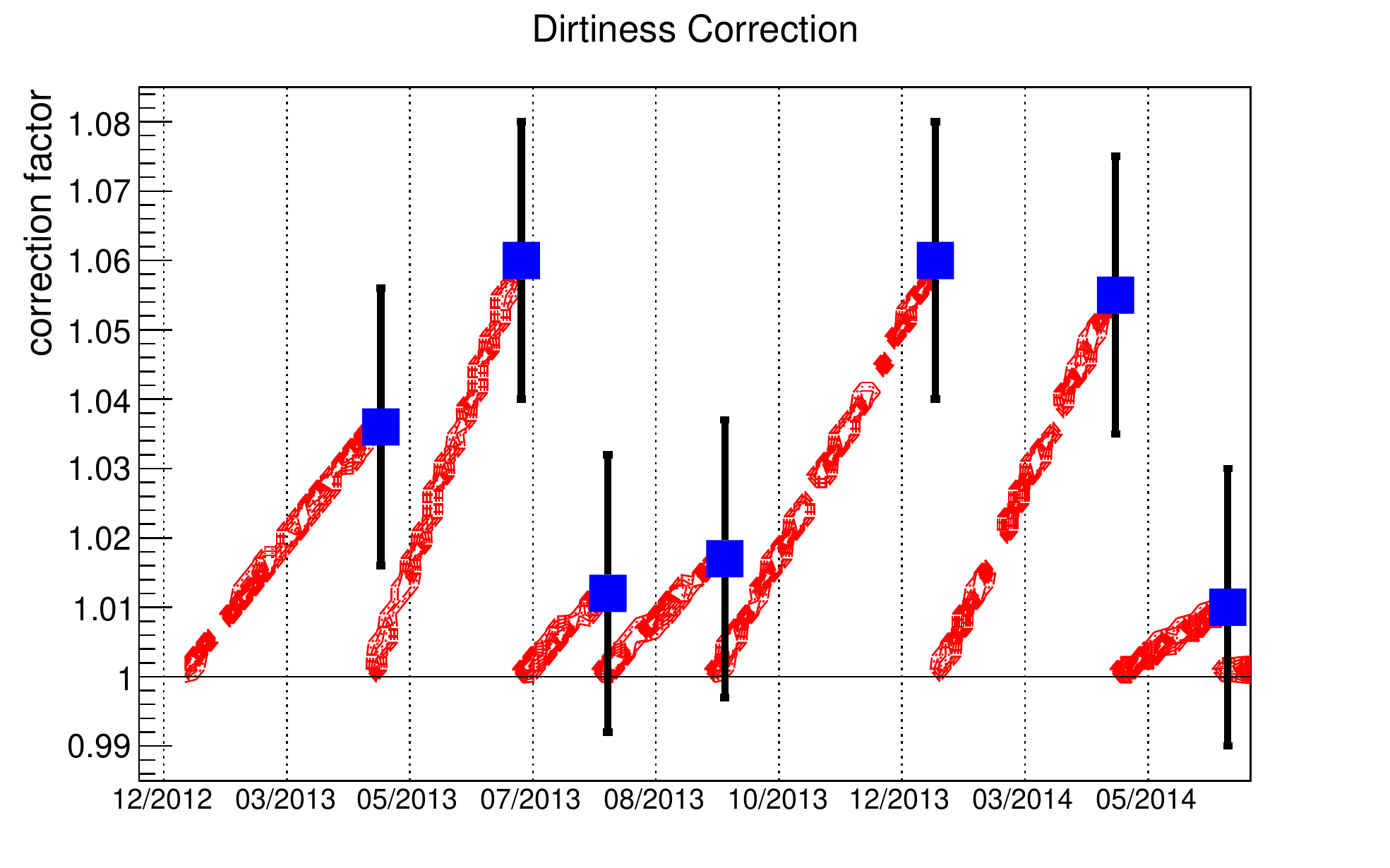}
\caption{Applied ``dirtiness'' corrections (red lines) to the light sensor data of one typical ATMOSCOPE, obtained from measurements (blue squares) as described Figure~\ref{fig:maluscalibrationexample}. 
%Each color corresponds to a different device at a different candidate site.
}
\label{fig:dirtiness}
\end{figure}

\begin{table}[h!]
\centering
\renewcommand{\arraystretch}{0.7}
{\small
\begin{tabular}{|l|c|c|p{4.2cm}|}
%%\begin{tabular}{|l|p{2.cm}|p{2.2cm}|p{4.2cm}|}
\hline
\multicolumn{4}{|c|}{Absolute Calibration LoNS Sensor} \\[0.2cm]
            & V-filter  & ATMOSCOPE    & Comments\\
            &           & ``B''-filter  & \\
\hline
Filter transmission &   1\% &  2\%  & spectral acceptance\\
Filter leakage      &   1\% &  3\%  & for typical dark sites \\
Angular acceptance  &   3\% &  3\%  & limitations of lab. \\
                    &       &       & measurements (spectral \\
                    &       &       & dependency, etc.)\\
Acceptance above 70$^\circ$ & $<$2\%    & $<$2\% & \\
Shadows ATMOSCOPE   & $<$1\% & $<$1\%  & \\
Dirtiness/deposits  & 2\%    & 2\%     & \\
Light leaks         & $<$1\% & $<$1\%  & \\
Lab. calibration    & 4\%    & 4\%     & \\
Entrance pupil lens & 2\%    & 2\%     & \\
Dark current subtraction &$<$2\% & $<$3\%&  \\ 
% SNR>30 (B) and >50 (V)
Temperature effects &$<$0.5\% & $<$0.5\% & typical temperature \\
                    &         &          & gradients\\ 
% 5 K --> change of 0.25 p.e./ns in worst case, ~ 100% of signal in worst case --> gradient of 1K ~20\%.
% maximum gradient: 20 K/h --> 5.5E-3 K/s --> <55E-3 K / (10 s measurement time): if dark current ~ signal, then d(signal)/signal <5.5 * 0.2\% = 1\%
\hline
Total               & $<$7\% & $<$8\% & \\
\hline
% 1%/K --> differences of CPPM (assume 20deg.C) and ref. site from other instrument (0deg C??)
%Temperature effects & 20\% & 20\% \\
\hline
\multicolumn{4}{|c|}{Relative inter-calibration LoNS Sensor between sites, after deployment} \\[0.2cm]
   & V-filter & ATMOSCOPE & Comments \\
   &          & ``B''-filter &  \\
\hline
Filter transmission & 0.5\%   & 1\% & spectral acceptance\\
Filter leakage      &   1\%   & 2\% & for typical dark sites \\
Angular acceptance  &   1\%   & 1\% & for typical dark sites \\
Shadows Surroundings &   1\%   & 1\% & \\
Interpolation Dirtiness & 2--8\%  & 2--8\% & on average $<$2\%, larger  \\
   &  &  & only for indiv. days\\
Inter-calibration lab. & 4\%    & 4\% & \\
Entrance pupil lens & $<$1\%    & $<$1\% & gluing, etc.\\
Dark current subtraction &$<$2\% & $<$3\%&  \\ 
Temperature effects &$<$1\% & $<$1\% & only for exceptionally \\
                    &       &        & strong temperature \\
                    &        &       & gradients\\ 
\hline
Total               & 5--9.5\% & 6--10\% & typically 5--6\% \\
\hline
%                    &           &        & have no dirtiness corrections (hence 8\%)\\
\end{tabular}
}
\caption{\label{tab:system} Compilation of systematic uncertainties of the ATMOSCOPE LoNS sensor.}
\end{table}

The sensitivity of the sensors is monitored on-site from time to time with a device as described above. In this case the main purpose is to
 measure the loss of transparency of the lens due to dust deposition: a measure of the LoNS sensor response is performed before and after cleaning the lens, and the difference is 
 attributed to dust deposits on the lens, the so-called ``dirtiness'' (see Figure~\ref{fig:maluscalibrationexample}). 
 Then, an approximation of the evolution of the loss of transparency along the time is performed by performing a linear interpolation
 between the day of the last cleaning and the current one.  
Typical correction factors range from zero to eight percent over periods of 
several months (see Figure~\ref{fig:dirtiness}). An absolute calibration of the LoNS sensor can be attempted either by comparison with other devices, or with the help 
of known stars passing across the field of view of the devices. The description of these methods is however outside the scope of this paper, and will be treated in a future publication. 

Table~\ref{tab:system} lists the residual systematic uncertainties after calibration, divided in absolute calibration uncertainties, valid for comparison of the results with those 
obtained from a different instrument -- such as external measurements by optical telescopes -- 
and relative uncertainties valid for the inter-comparison of sites. The totals have been calculated by adding the individual items in quadrature.

\section{The Sky Quality Meter}

\begin{figure}
                \begin{center}
                        \includegraphics[height=0.5\columnwidth]{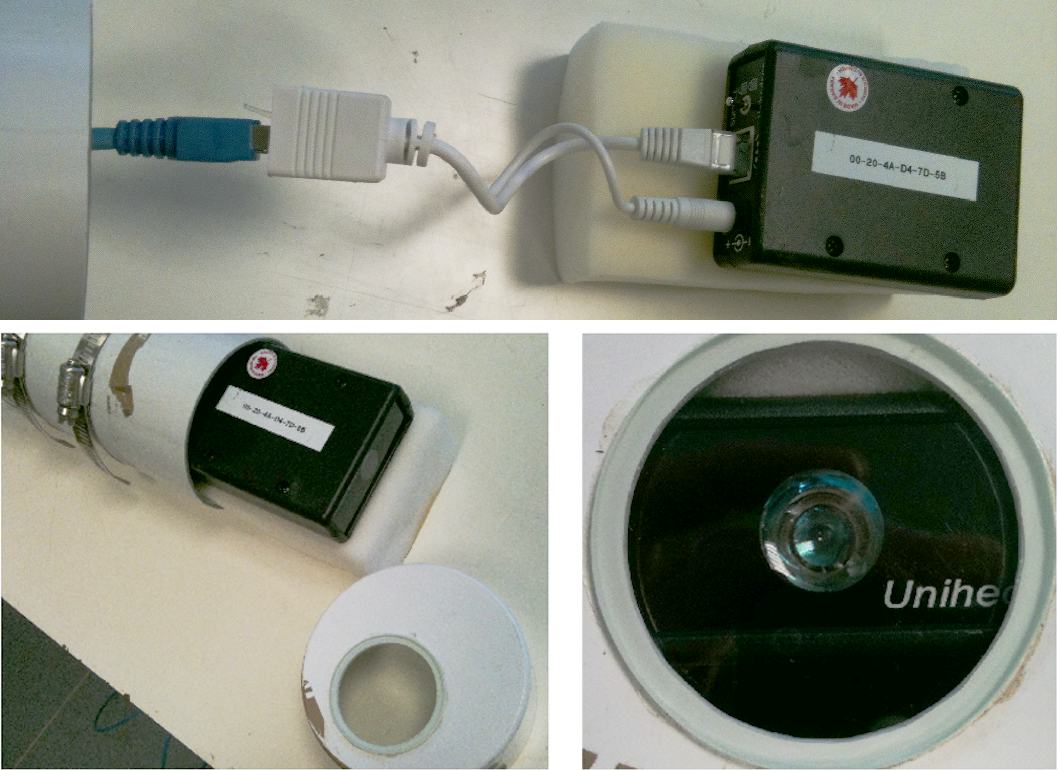}
                \end{center}
                \caption{Details of the housing of the SQM: The SQM is powered using the Ethernet cable (Power over Ethernet, 5V) and the readout is performed each minute across the
                 Ethernet by a simple script. The glass housing protects the SQM against the rain.}
                \label{fig:sqm}
        \end{figure}

A commercial LoNS measurement device, a Sky Quality Meter SQM-LE, from Unihedron, is also part of the ATMOSCOPE 
(see Figure~\ref{fig:sqm}). While the SQMs are not as accurate as the LoNS,
we have decided to use them for the following reasons: the SQMs are extremely robust, and can work for extended
periods of time with no maintenance. Moreover they are the standard tools to measure the night sky background
providing a check of the results as well as easy comparison to many sites worldwide.

The precision is $\pm$0.1 mag/arcsec$^2$ according to the provider, and the operating temperature range is $-40$ to $85^\circ$C. The SQM is housed in a protective
tube with a glass window, and operates continuously, with one measurement being performed each minute. 
% The angular acceptance of the SQM has been measured in the laboratory, and is compared to the one of the LoNS device in Figure~\ref{fig:sqmacceptance}. 
The angular acceptance of the SQM is 20$\deg$ FWHM.
SQMs are known to have considerably broader spectral acceptance than 
through a B and V-filter alone (recall Figure~\ref{fig:filters}). Cinzano~\citep{cinzano} has shown that the broader acceptance may result in typical LoNS measurements taken with the SQM 
differing from standard V-band measurements by up to 0.25~mag/arcsecond$^2$.
       % \begin{figure}
       %         \begin{center}
       %                 \includegraphics[height=0.5\columnwidth]{img/fig_acceptancelons-comparedsqm.png}
        %        \end{center}
        %        \caption{Acceptance of the SQM compared to the one of the LoNS sensor.}
        %        \label{fig:sqmacceptance}
        %\end{figure}
        
\begin{figure}[h!]
\centering
\vspace{-2.5cm}
\includegraphics[width=0.8\columnwidth]{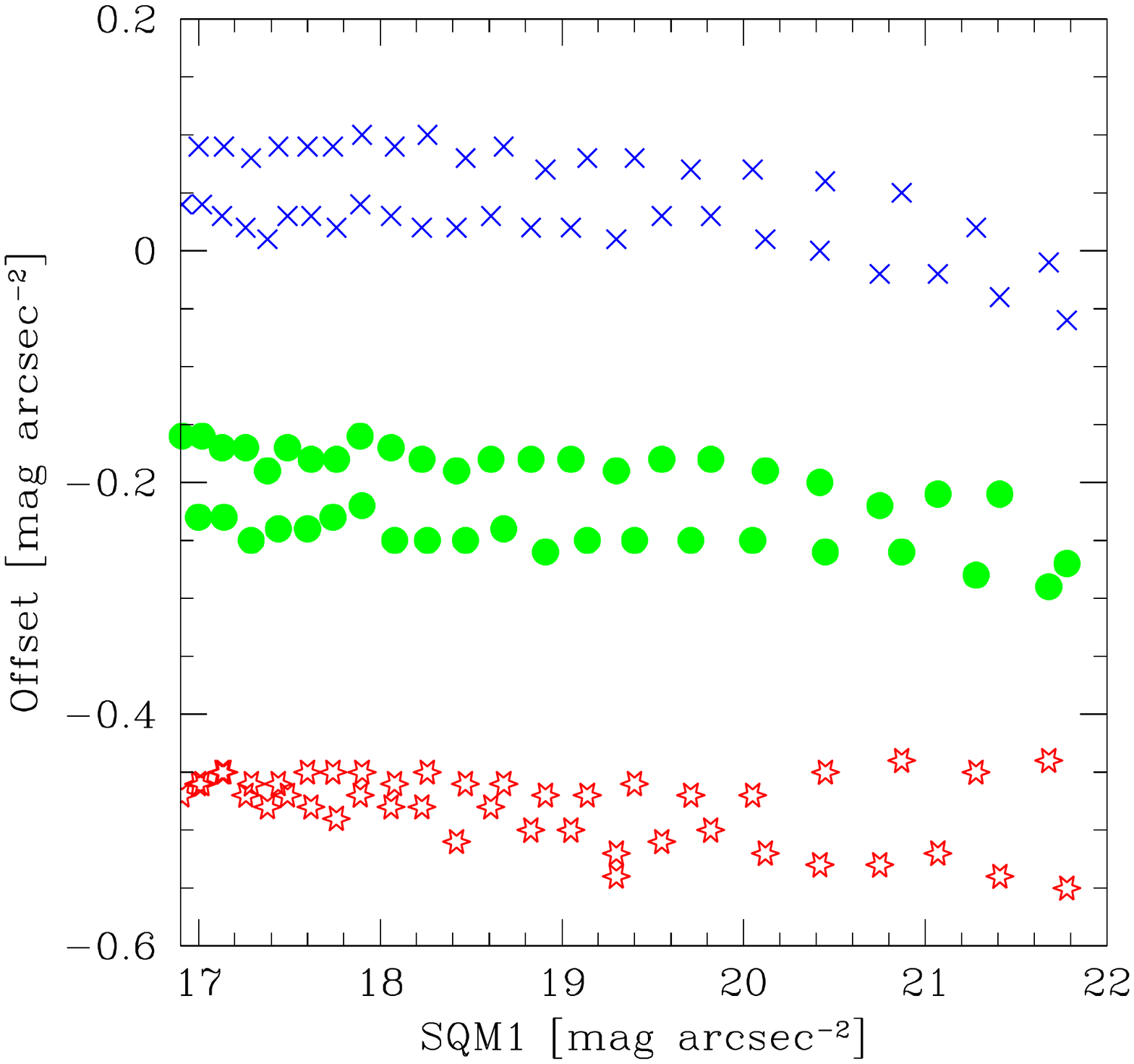}
\vspace{-3cm}
\caption{An example of the cross-calibration measurement of the SQMs. 
Shown are the differences between the reference SQM and three other units as a function of the reference SQM value.}
\label{sqm:cross}
\end{figure}

The actual accuracy of the SQM was found to be worse than the specifications.
To cross-calibrate the SQMs, they were placed side by side in a
controlled low light environment, as the light intensity was varied. We present 
an example of the results in Figure~\ref{sqm:cross}. The Figure shows the measured
differences between three SQMs and one that was chosen as reference.
The results show substantial offsets of up to  $\pm$0.4 mag/arcsec$^2$ between 
various units. However, for each unit individually, the spread between the measurements is smaller than the 
$\pm$0.1~mag/arcsec$^2$ listed in the specifications. The offsets measured 
in the laboratory were later used in the data analysis. 

Table~\ref{tab:instr_lons} lists the most important technical details of the two devices employed  to measure the LoNS.

\afterpage{
\begin{table}\footnotesize
\centering
\begin{tabular}{ c | c }
\hline
instruments & technical properties \\
\hline \hline
	  LoNS sensor & 
	  \begin{tabular}{ c c }
	    \begin{tabular}{ c } acceptance \\ intercalib. precision \end{tabular}
	    \begin{tabular}{ c } $0.375 \pm \np[sr]{0.005}$ \\ $\np[mag/arcsec^{2}]{0.06}$  \end{tabular}
	  \end{tabular} \\
	  \hline
	  SQM & 
	  \begin{tabular}{ c c }
	    \begin{tabular}{ c } 
							acceptance\footnotemark \\ %resolution \\ 
              intercalib. precision  \\
							temperature range
			\end{tabular}
	    \begin{tabular}{ c } 
							$0.17 \pm \np[sr]{0.02}$ \\ % $\np[mag/arcsec^{2}]{0.1}$ \\ 
              $\np[mag/arcsec^{2}]{0.17}$ \\
							$-40 \ \mathrm{to} \ \np[\deg C]{85}$
			\end{tabular}
	  \end{tabular} \\
	  \hline
	  
\end{tabular}
\caption{Summary of different instruments to measure the LoNS and the most important technical details.}
\label{tab:instr_lons}
\end{table}
\footnotetext{Official manufacturer claim: $20 \deg$ FWHM, corresponding to $0.17 \pm \np[sr]{0.02}$ according to the shape of the efficiency curve provided by Unihedron.}
}

\section{Control computer and software}
	
	\begin{figure}
		\begin{center}
			\includegraphics[height=0.55\columnwidth]{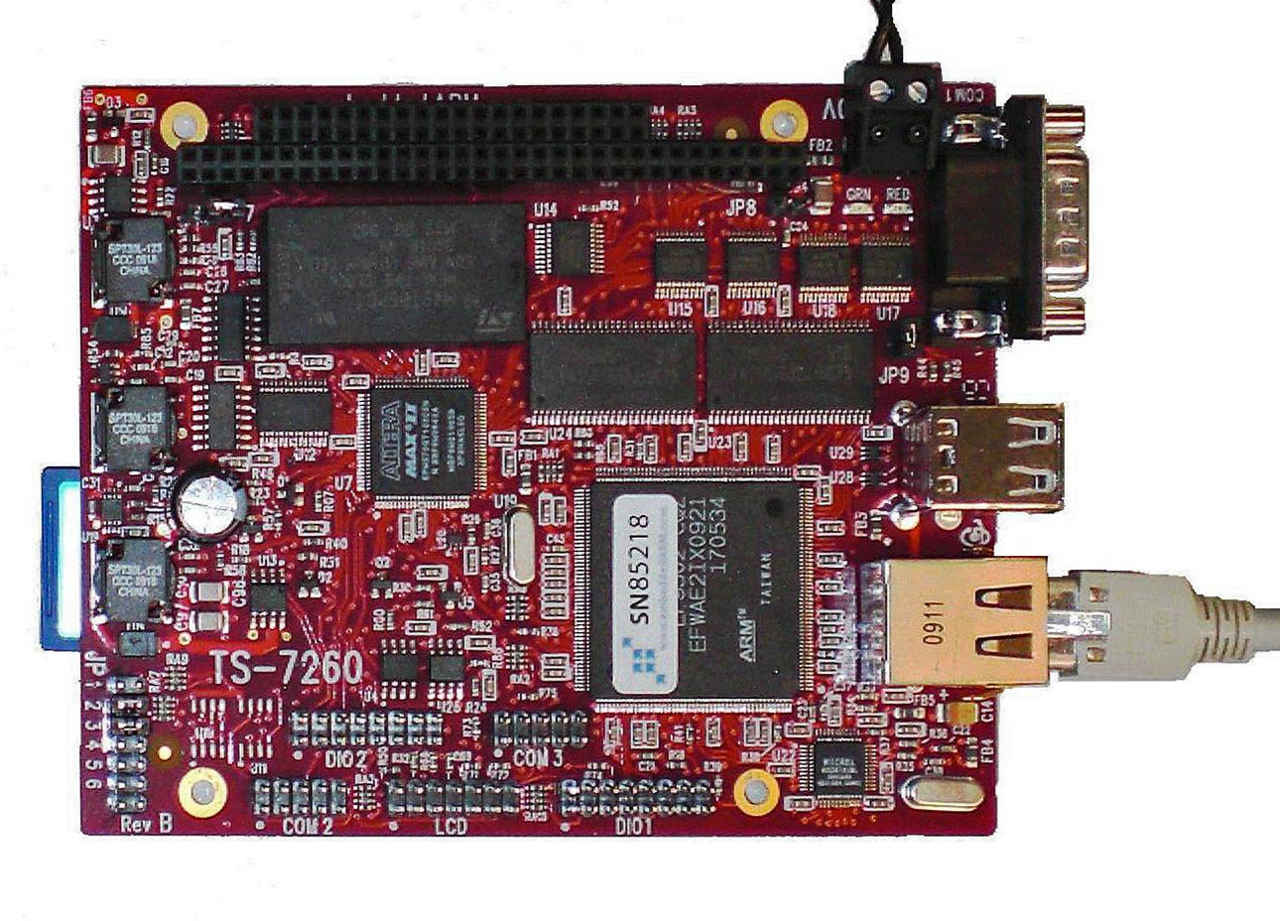}
		\end{center}
		\caption{TS-7260 single board computer, used for the operation and readout of the ATMOSCOPE.}
		\label{fig:ts7260}
	\end{figure}

\begin{figure}
		\begin{center}
			\includegraphics[height=0.55\columnwidth]{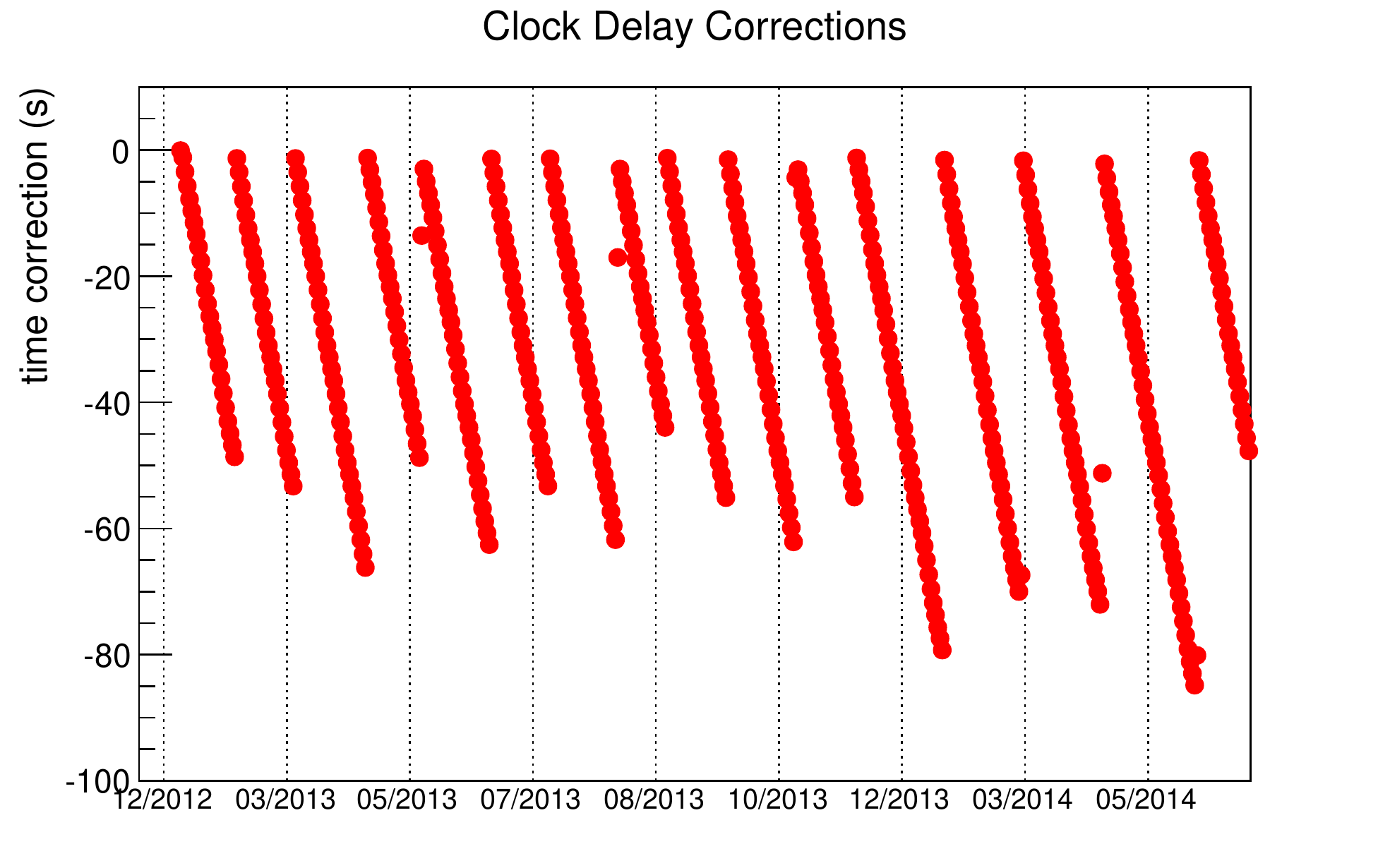}
		\end{center}
		\caption{Internal clock delays, as estimated from one typical ATMOSCOPE. The jumps correspond to those moments when the clock was synchronized.}
		\label{fig:delays}
	\end{figure}

As control computer we use a TS-7260 single board computer (see Figure~\ref{fig:ts7260}). 
It uses an ARM9 CPU, running at $\np[MHz]{200}$, has an SD-card slot that can be equipped with up to 2~GB memory and a battery-supported real time clock.
The overall power consumption is less than $\np[W]{1}$. In addition, it has three serial ports, one of which is used to read out the data from the weather station. 
There are also 30 independently usable I/O lines that are used for the I$^2$C communication and the motor control, 
and a watch dog timer, to restart the board in case it hangs. 
The internal clocks of the ATMOSCOPEs show considerable drifts, advancing by on average 20~$\mu$s/s, with a spread of 5~$\mu$s/s between different clocks (see Figure~\ref{fig:delays}). 
%The clocks are synchronized with a GPS reference from time to time, the parts between synchronizations are corrected assuming a linear drift with time. 
Monthly synchronizations of the clocks have been made since autumn 2012. 
This has reduced the clock drift corrections to always less than one minute, the data between synchronizations being corrected assuming a linear drift with time.
	
The computer runs a Debian Linux operating system that is installed on and booted directly from SD-card. 
This makes it possible to compile the software on board. 
Setting up the system for a second or third station means just changing the boot procedure to SD-boot and copying the SD card. 
The data are stored on a USB key connected to one of the 2 available USB ports on the board.  
The data files are uploaded each hour on a data server located in Warsaw, using the \textit{rsync} linux utility. For this purpose, all ATMOSCOPEs are connected to the network 
via 3G connections or wireless links when a nearby observatory is reachable. This allows also for an hourly monitoring of ATMOSCOPE measurements,
as well as remote checks of the system and the monthly tuning of clocks mentioned above. The network link is
achieved using a router (3G or Wireless) and a switch serving a local network loop to which the TS-7260 board is connected, 
as well as the SQM, the All Sky Camera PC, and, on some sites, auxiliary devices.\\

All programs are written in \textit{C} and called by entries in the \textit{crontab}. %he periodicity of LoNS sensor measurements is one minute.
After the measurement process is launched, the filter wheel is still in the dark frame position, as set after the last measurement. 
First the temperature sensors and the humidity monitor, attached to the ADC, are read, and the physical values are calculated from the voltage values. 
Then the three ADC channels, assigned to the LoNS sensor amplification stages, are read out 1000 times each and averaged to increase the precision. 
The statistical uncertainties are also calculated to get an estimate of the fluctuations. 
All voltages are converted back to the corresponding currents using the appropriate amplification factors. 
During daytime, too much light reaches the PIN diode for the measurements to be useful. 
Therefore, if the voltage value in the first amplification stage exceeds $\np[mV]{100}$, the wheel does not turn at all so that the PIN diode is protected. 
If it is dark enough outside, the four other filters are selected in sequence and 
the voltage measurement procedure described above is repeated each time.
The weather station data taken simultaneously are added to the measurement results of the LoNS device. 
%The values are all transmitted in one string. 
Finally, all values and errors, together with the system date and the Unix time stamp,
gathered in one single string, are appended to a file.\\

The SQM data are recorded each minute by sending a simple reading request over the local network loop; 
they are stored in a separate file. Each record consists of the system date, the magnitude per square arc second,
the temperature at the level of the light sensor, and few auxiliary parameters.

\section{Conclusion}

Ten ATMOSCOPEs (Autonomous Tool for Measuring Observatory Site COnditions PrEcisely) have been built 
and installed at 9 CTA candidate sites in the Northern and Southern hemisphere, while one reference unit stayed in the laboratory. 
Each apparatus allows the measurement of weather parameters by means of a commercial weather station measuring
wind, temperature, humidity, atmospheric pressure and cloud altitude, and allows the measurement of the Light of the Night Sky (LoNS) by means of two sensors: 
a commercial one (Sky Quality Meter) and one specific sensor designed and built at the MPI Munich. 
The latter is based on a large-area calibrated PIN-photodiode and two filters, in the V~band and a customized ``ATMOSCOPE-B'' band, which matches the 
spectral acceptance of super-bialkali photomultipliers. The ATMOSCOPEs are complemented by All-Sky-Cameras to derive a cloud fraction parameter during the night.

On all sites the weather station, initially held by the ATMOSCOPE structure at 2.5~m above the ground, has been relocated to the top of a mast in 10 m height for being less sensitive to aerodynamic ground effects caused by surface roughness and vegetation. 
The optical devices, assisted by the weather measurements, have performed a precise LoNS estimate during the clear nights, measuring the light of the sky each minute
over periods of up to 2.5 years, depending on the site. By means of wireless or 3G communication, depending on the presence of neighboring observatories,
data have been retrieved hourly to allow for online monitoring. 

\begin{figure}
                \begin{center}
                        %\includegraphics[width=0.99\columnwidth]{img/fig_example_lons_sqm.png}
			% here you can choose with or without galactic latitude (of the zenith direction)
			\includegraphics[width=0.99\columnwidth]{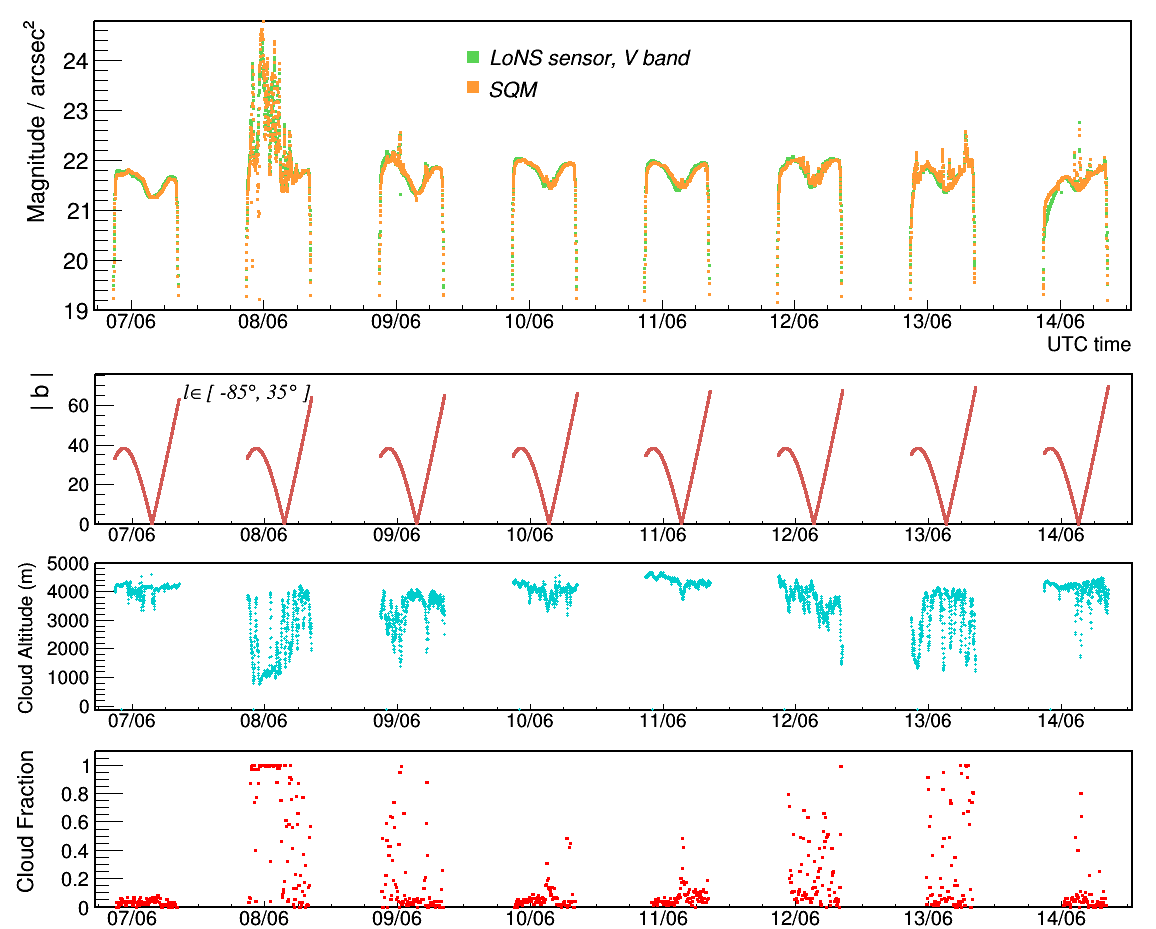}
                \end{center}
                \caption{Example of data provided by the LoNS sensor, the SQM (top) and the cloud sensor and All-Sky-Camera (bottom panels) running together. 
The Galactic latitude observed by the sensors in the zenith direction is shown on the second panel.}
                \label{fig:example_lons_sqm}
        \end{figure}

The data provided by the devices equipping the ATMOSCOPEs have been analyzed and used to give an evaluation
of each site. The analysis methods applied and results obtained will be presented in another publication.
Figure~\ref{fig:example_lons_sqm} shows an example of raw data obtained by the two optical sensors taking data simultaneously over 8 days, with the cloud altitude given by the weather station
and the cloud fraction measured with the All-Sky-Camera.
The correlation between the measurements can be seen directly: 
a darkening of the sky seen by the optical devices (larger mag/arcsec$^2$) correlates with the presence of clouds,  
indicated by the decreasing cloud altitude measured by the cloud sensor, the increasing cloud fraction (here within $20^\circ$ of the zenith direction) given by the All-Sky-Camera 
and large variability of the measured light. 
The V-shape of the LoNS measurement is due to the Galactic plane passing across the field of view of the optical devices. 
The last night displayed on the top panel shows the influence of the Moon which transits above the
horizon at the beginning of the night.

\FloatBarrier				
\section{Acknowledgments}

We gratefully acknowledge financial support from
the following agencies and organizations:

Ministerio de Ciencia, Tecnolog\'ia e Innovaci\'on Productiva (MinCyT),
Comisi\'on Nacional de Energ\'ia At\'omica (CNEA), Consejo Nacional
de Investigaciones Cient\'ificas y T\'ecnicas (CONICET), Argentina;
State Committee of Science of Armenia, Armenia;
Conselho Nacional de Desenvolvimento Cient\'{i}fico e Tecnol\'{o}gico (CNPq),
Funda\c{c}\~{a}o de Amparo \`{a} Pesquisa do Estado do Rio de Janeiro (FAPERJ),
Funda\c{c}\~{a}o de Amparo \`{a} Pesquisa do Estado de S\~{a}o Paulo (FAPESP), Brasil;
Croatian Science Foundation, Croatia; 
Ministry of Education, Youth and Sports, MEYS LE13012, 7AMB12AR013, Czech Republic;
Ministry of Higher Education and Research, CNRS-INSU and CNRS-IN2P3, CEA-Irfu, ANR,
Regional Council Ile de France, Labex ENIGMASS, OSUG2020 and OCEVU, France;
Max Planck Society, BMBF, DESY, Helmholtz Association, Germany;
Department of Atomic Energy, Department of Science and Technology, India;
Istituto Nazionale di Astrofisica (INAF), MIUR, Italy;
ICRR, University of Tokyo, JSPS, Japan;
Netherlands Research School for Astronomy (NOVA),
Netherlands Organization for Scientific Research (NWO), Netherlands;
The Bergen Research Foundation, Norway;
Ministry of Science and Higher Education, the National Centre for Research and 
Development and the National Science Centre, Poland;
MINECO support through the National R+D+I, CDTI funding plans and the CPAN and MultiDark
Consolider-Ingenio 2010 programme, Spain;
Swedish Research Council, Royal Swedish Academy of Sciences, Sweden;
Swiss National Science Foundation (SNSF), Ernest Boninchi Foundation, Switzerland;
Durham University, Leverhulme Trust, Liverpool University, University of Leicester,
University of Oxford, Royal Society, Science and Technologies Facilities Council, UK;
U.S. National Science Foundation, U.S. Department of Energy,
Argonne National Laboratory, Barnard College, University of California,
University of Chicago, Columbia University, Georgia Institute of Technology,
Institute for Nuclear and Particle Astrophysics (INPAC-MRPI program),
Iowa State University, Washington University McDonnell Center for the Space Sciences, USA.

The research leading to these results has received funding from the
European Union's Seventh Framework Programme (FP7/2007-2013) under grant
agreement nr. 262053 and is supported by the Ministry of Education, Youth and Sports of the
Czech Republic within the projects LE13012 and LG14019.

\noindent Tomasz Bulik is grateful for the support from MNISW through grant 2011/01/M/ST9/01891.

\noindent We thank Paula Chadwick for the exhaustive English language corrections. 

Finally we would like to acknowledge the influence on the project that came
from the late Prof. Eckart Lorentz. He has been the author of the original
idea to build the ATMOSCOPEs, and he has guided and mentored
us during the prototyping and construction of the project.
We are all very grateful for the opportunity to work with him
and we will miss his advice.

\clearpage

\bibliographystyle{JHEP}
%\bibliographystyle{elsarticle-harv}
%\bibliography{bibliography}

\providecommand{\href}[2]{#2}\begingroup\raggedright\endgroup

%\bsp
%\label{lastpage}

\end{document}